\documentclass[journal]{IEEEtran}

\usepackage[usenames,dvipsnames,svgnames,table]{xcolor}

\usepackage[printonlyused]{acronym}

\acrodef{DPDK}{Data Plane Development Kit}
\acrodef{vRAN}{virtualized Radio Access Network}
\acrodef{ILP}{Integer Linear Program}
\acrodef{VM}{Virtual Machine}
\acrodef{C-RAN}{Cloud Radio Access Network}
\acrodef{RAN}{Radio Access Network}
\acrodef{UPnP}{Universal Plug and Play}
\acrodef{NAT}{Network Address Translation}
\acrodef{DHCP}{Dynamic Host Configuration Protocol}
\acrodef{ISG}{Industry Standards Group}
\acrodef{3GPP}{3rd Generation Partnership Project}
\acrodef{EPC}{Evolved Packet Core}
\acrodef{CPE}{Customer Premises Equipment}
\acrodef{LTE}{Long Term Evolution}
\acrodef{HA}{Hardware Acceleration}
\acrodef{ETSI}{European Telecommunications Standards Institute}
\acrodef{InP}{Infrastructure Provider}
\acrodef{SDN}{Software Defined Networking}
\acrodef{VN}{Virtual Network}
\acrodef{QoE}{Quality-of-Experience}
\acrodef{QoS}{Quality-of-Service}
\acrodef{VNF}{Virtual Network Function}
\acrodef{NFVI}{Network Function Virtualization Infrastructure}
\acrodef{NF}{Network Function}
\acrodef{SN}{Substrate Network}
\acrodef{vCPE}{virtual Customer Premises Equipment}
\acrodef{vEPC}{virtual Evolved Packet Core}
\acrodef{VNE}{Virtual Network Embedding}
\acrodef{SP}{Service Provider}
\acrodef{ARPU}{Average Revenue Per User}
\acrodef{MANO}{Management and Orchestration}
\acrodef{TSP}{Telecommunication Service Provider}
\acrodef{NFV}{Network Function Virtualization}
\acrodef{CAPEX}{Capital Expenses}
\acrodef{OPEX}{Operating Expenses}
\acrodef{DPI}{Deep Packet Inspection}
\acrodef{ONF}{Open Network Foundation}

\ifCLASSINFOpdf
   \usepackage[pdftex]{graphicx}

   \graphicspath{{../pdf/}{figures/}}
   \DeclareGraphicsExtensions{.pdf,.jpeg,.png,.jpg}
\else
\fi
\usepackage[cmex10]{amsmath}
\usepackage{algorithmic}
\usepackage{algorithm}
\usepackage{array}
\newcolumntype{L}[1]{>{\raggedright\let\newline\\\arraybackslash\hspace{0pt}}m{#1}}
\newcolumntype{C}[1]{>{\centering\let\newline\\\arraybackslash\hspace{0pt}}m{#1}}
\newcolumntype{R}[1]{>{\raggedleft\let\newline\\\arraybackslash\hspace{0pt}}m{#1}}

\usepackage{multirow}

\usepackage{url}

\usepackage[table]{xcolor}

\makeatletter
\def\ps@IEEEtitlepagestyle{%
  \def\@oddhead{\strut\hfill 
  
\begin{tabular}{m{17.5cm}}
          \centering
          \footnotesize{This is the author's version of an article that has been published in IEEE Communications Surveys and Tutorials. Changes were made to this version by the publisher prior to publication. The final version of record is available at {\color{blue}http://dx.doi.org/10.1109/COMST.2015.2477041}}
          \tabularnewline
          \end{tabular}%
\hfill\strut}%
  \def\@oddfoot{\mycopyrightnotice}%
  \def\@evenfoot{}%
}

\def\mycopyrightnotice{%
  {
\begin{tabular}{m{19.0cm}}
          \centering
          \footnotesize{Copyright (c) 2015 IEEE. Personal use is permitted. For any other purposes, permission must be obtained from the IEEE by emailing pubs-permissions@ieee.org.}
          \tabularnewline
          \end{tabular}%
  
  \hfill}
  \gdef\mycopyrightnotice{}
}

\begin{document}

%
\title{Network Function Virtualization: State-of-the-art and Research Challenges}
%
%
%


\author{Rashid~Mijumbi, Joan~Serrat, Juan-Luis~Gorricho, Niels~Bouten, Filip~De~Turck, Raouf~Boutaba

\thanks{R. Mijumbi, J. Serrat and J.L. Gorricho are with the Network Engineering Department, Universitat Polit\`{e}cnica de Catalunya, 08034 Barcelona, Spain.}
\thanks{N. Bouten and F. De Turck are with Department of Information Technology, Ghent University - iMinds, B-9050 Ghent, Belgium.}
\thanks{R. Boutaba is with the D.R. Cheriton School of Computer Science, University of Waterloo, Waterloo, Ontario, N2L 3G1, Canada.}

}


%
%

\markboth{IEEE Communications Surveys \& Tutorials}
{Shell \MakeLowercase{\textit{et al.}}: Bare Demo of IEEEtran.cls for Journals}
%



\maketitle

\begin{abstract}
\ac{NFV} has drawn significant attention from both industry and academia as an important shift in telecommunication service provisioning. By decoupling \acp{NF} from the physical devices on which they run, NFV has the potential to lead to significant reductions in \ac{OPEX} and \ac{CAPEX} and facilitate the deployment of new services with increased agility and faster time-to-value. The NFV paradigm is still in its infancy and there is a large spectrum of opportunities for the research community to develop new architectures, systems and applications, and to evaluate alternatives and trade-offs in developing technologies for its successful deployment. In this paper, after discussing NFV and its relationship with complementary fields of \ac{SDN} and cloud computing, we survey the state-of-the-art in NFV, and identify promising research directions in this area. We also overview key NFV projects, standardization efforts, early implementations, use cases and commercial products.
\end{abstract}


%
\IEEEpeerreviewmaketitle

\section{Introduction}

Service provision within the telecommunications industry has traditionally been based on network operators deploying physical proprietary devices and equipment for each function that is part of a given service. In addition, service components have strict chaining and/or ordering that must be reflected in the network topology and in the localization of service elements. These, coupled with requirements for high quality, stability and stringent protocol adherence, have led to long product cycles, very low service agility and heavy dependence on specialized hardware. 

However, the requirements by users for more diverse and new (short-lived) services with high data rates continue to increase. Therefore, \acp{TSP} must correspondingly and continuously purchase, store and operate new physical equipment. This does not only require high and rapidly changing skills for technicians operating and managing this equipment, but also requires dense deployments of network equipment such as base stations. All these lead to high \ac{CAPEX} and \ac{OPEX} for TSPs \cite{JunWuZhi15, ChinaMobile}.

Moreover, even with these high customer demands, the resulting increase in capital and operational costs cannot be translated in higher subscription fees, since TSPs have learned that due to the high competition, both among themselves and from services being provided over-the-top on their data channels, increasing prices only leads to customer churn. Therefore, TSPs have been forced to find ways of building more dynamic and service-aware networks with the objective of reducing product cycles, operating \& capital expenses and improving service agility. 

\ac{NFV} \cite{HanBoGo15, Guerzoni12} has been proposed as a way to address these challenges by leveraging virtualization technology to offer a new way to design, deploy and manage networking services. The main idea of NFV is the decoupling of physical network equipment from the functions that run on them. This means that a network function - such as a firewall - can be dispatched to a TSP as an instance of plain software. This allows for the consolidation of many network equipment types onto high volume servers, switches and storage, which could be located in data centers, distributed network nodes and at end user premises. This way, a given service can be decomposed into a set of \acp{VNF}, which could then be implemented in software running on one or more industry standard physical servers. The VNFs may then be relocated and instantiated at different network locations (e.g., aimed at introduction of a service targeting customers in a given geographical location) without necessarily requiring the purchase and installation of new hardware.
\begin{figure*}[t]
\begin{minipage}{.5\textwidth}
\resizebox{.99\textwidth}{!}
{\includegraphics{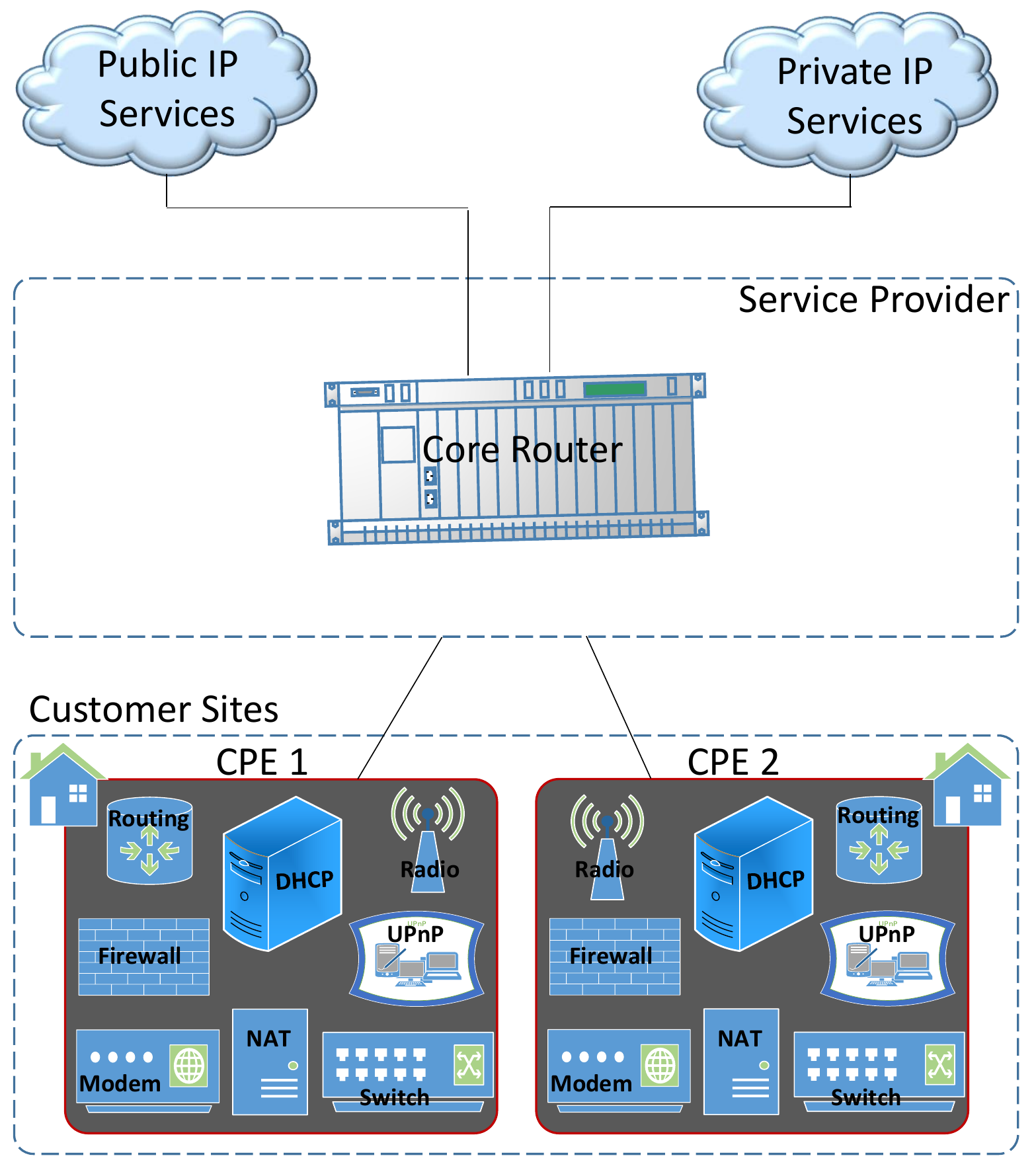}}
  \caption{Traditional CPE Implementations}\label{current}
\end{minipage}
\begin{minipage}{.5\textwidth}
\resizebox{.99\textwidth}{!}
{\includegraphics{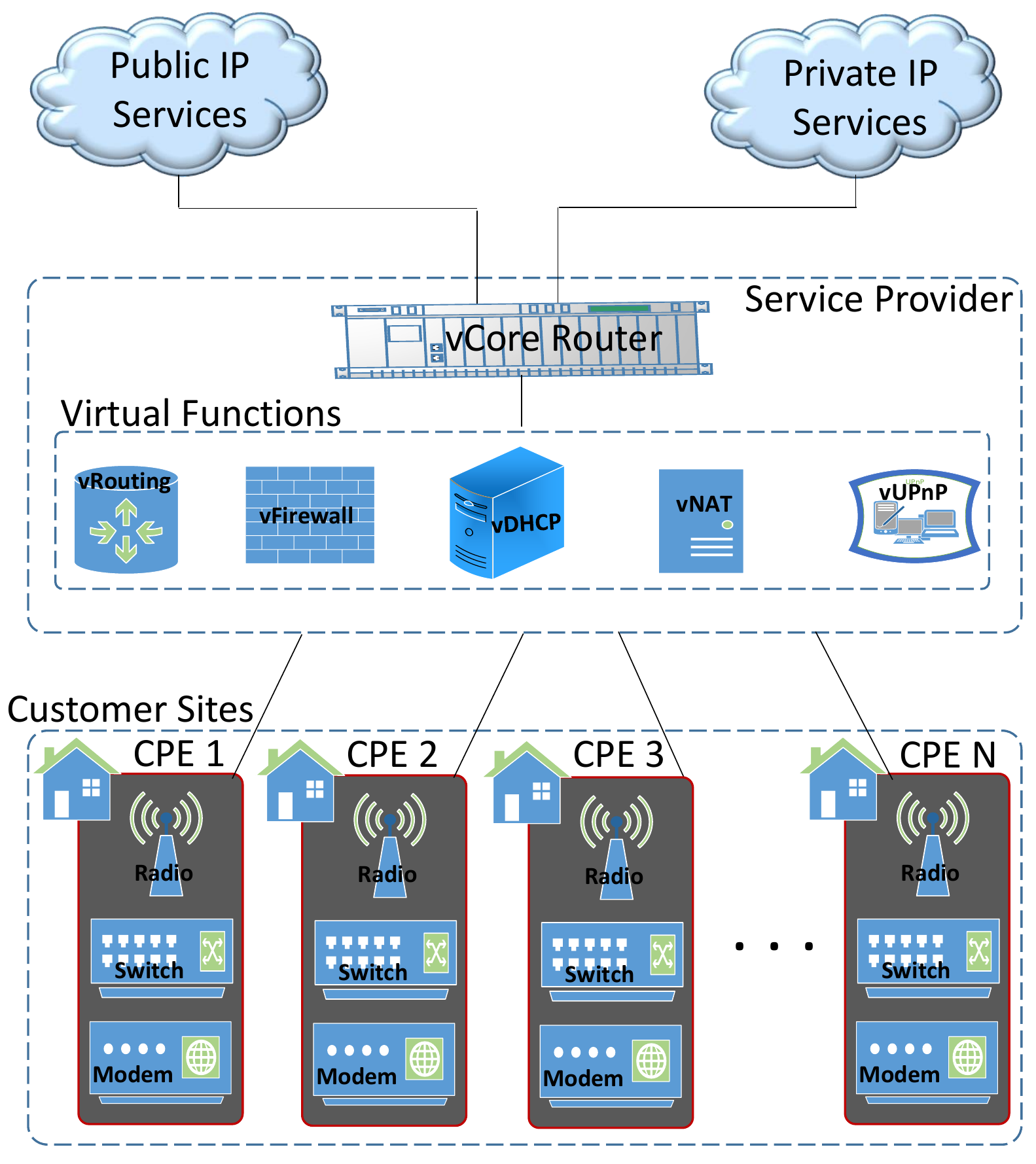}}
  \caption{Possible CPE Implementation with NFV}
  \label{withNFV}
\end{minipage}
\end{figure*}

\ac{NFV} promises TSPs with more flexibility to further open up their network capabilities and services to users and
other services, and the ability to deploy or support new network services faster and cheaper so as to realize better
service agility. To achieve these benefits, NFV paves the way to a number of differences in the way network service provisioning is realized in comparison to current practice. In summary, these differences are as follows \cite{ETSINFV002}:

\textit{Decoupling software from hardware}. As the network element is no longer a composition of integrated hardware
and software entities, the evolution of both are independent of each other. This allows separate development timelines and maintenance for software and hardware.

\textit{Flexible network function deployment}. The detachment of software from hardware helps reassign and share the
infrastructure resources, thus together, hardware and software, can perform different functions at various
times. This helps network operators deploy new network services faster over the same physical platform.  Therefore, components can be instantiated at any NFV-enabled device in the network and their connections can be set up in a flexible way.

\textit{Dynamic scaling}. The decoupling of the functionality of the network function into instantiable software
components provides greater flexibility to scale the actual VNF performance in a more dynamic way and with
finer granularity, for instance, according to the actual traffic for which the network operator needs to provision
capacity.

It is worth remarking that the general concept of decoupling \acp{NF} from dedicated hardware does not necessarily require virtualization of resources. This means that \acp{TSP} could still purchase or develop software (\acp{NF}) and run it on physical machines. The difference is that these \acp{NF} would have to be able to run on commodity servers. However, the gains (such as flexibility, dynamic resource scaling, energy efficiency) anticipated from running these functions on virtualized resources are very strong selling points of NFV. Needless to mention, it is also possible to have hybrid scenarios where functions running on virtualized resources co-exist with those running on physical resources. Such hybrid scenarios may be important in the transition towards NFV.

\subsection{History of Network Function Virtualization}
The concept and collaborative work on NFV was born in October 2012 when a number of the world's leading TSPs jointly authored a white paper \cite{Guerzoni12} calling for industrial and research action. In November 2012 seven of these operators (AT\&T, BT, Deutsche Telekom, Orange, Telecom Italia, Telefonica and Verizon) selected the European Telecommunications Standards Institute (ETSI)\cite{ETSI} to be the home of the Industry Specification Group for NFV (ETSI ISG NFV)\footnote{In the rest of this paper, the acronyms ETSI and ETSI ISG NFV are used synonymously.}. Now, more than two years later, a large community of experts are working intensely to develop the required standards for NFV as well as sharing their experiences of its development and early implementation. The membership of ETSI has grown to over 245 individual companies including 37 of the world's major service providers as well as representatives from both telecoms and IT vendors \cite{ETSI}. ETSI has successfully completed Phase 1 of its work with the publication of 11 ETSI Group Specifications \cite{ETSIDOCS}. These specifications build on the first release of ETSI documents published in October 2013 and include an infrastructure overview, updated architectural framework, and descriptions of the compute, hypervisor and network domains of the infrastructure. They also cover \ac{MANO}, security and trust, resilience and service quality metrics. 

Since ETSI is not a standards body, its aim is to produce requirements and potential specifications that TSPs and equipment vendors can adapt for their individual environments, and which may be developed by an appropriate standards development organization (SDO). However, since standards bodies such as the 3GPP \cite{3GPP} are in liaison with the ETSI, we can expect these proposals will be generally accepted and enforced as standards. 3GPP's Telecom Management working group (SA5) is also studying the management of virtualized 3GPP network functions.

\subsection{NFV Examples}
The ETSI has proposed a number of use cases for NFV \cite{ETSIUseCases}. In this subsection, we will explain how NFV may be applied to \ac{CPE}, and to an \ac{EPC} network.
\subsubsection{Customer Premises Equipment (CPE)}
In Figures \ref{current} and \ref{withNFV}, we use an example of a \ac{CPE} to illustrate the economies of scale that may be achieved by NFV. Fig. \ref{current} shows a typical (current) implementation of a CPE which is made up of the functions: \ac{DHCP}, \ac{NAT}, routing, \ac{UPnP}, Firewall, Modem, radio and switching. In this example, a single service (the CPE) is made up of eight functions. These functions may have precedence requirements. For example, if the functions are part of a service chain\footnote{The chain of functions that make up a service for which the connectivity order is important is know as VNF Forwarding Graph (VNFFG) \cite{ETSIUseCases}. In addition to sequencing requirements, the links in a VNFFG may split (i.e. from one function, packets could take one of many paths which lead to similar functionality), or may join.}, it may be required to perform firewall functions before NAT. Currently, it is necessary to have these functions in a physical device located at the premises of each of the customers 1 and 2. With such an implementation, if there is a need to make changes to the CPE, say, by adding, removing or updating a function, it may be necessary for a technician from the ISP to individually talk to or go to each of the customers. It may even require a complete change of the device in case of additions. This is not only expensive (operationally) for the ISPs, but also for the customers.

In Figure \ref{withNFV}, we show a possible implementation based on NFV in which some of the functions of the CPE are transferred to a shared infrastructure at the ISP, which could also be a data center. This makes the changes described above easier since, for example, updating the DHCP for all customers would only involve changes at the ISP. In the same way, adding another function such as parental controls for all or a subset of customers can be done at once. In addition to saving on operational costs for the ISP, this potentially leads to cheaper CPEs if considered on a large scale.

\subsubsection{Evolved Packet Core}
\begin{figure}[t]
\centering
\includegraphics[width=8.55cm, height=8.20cm]{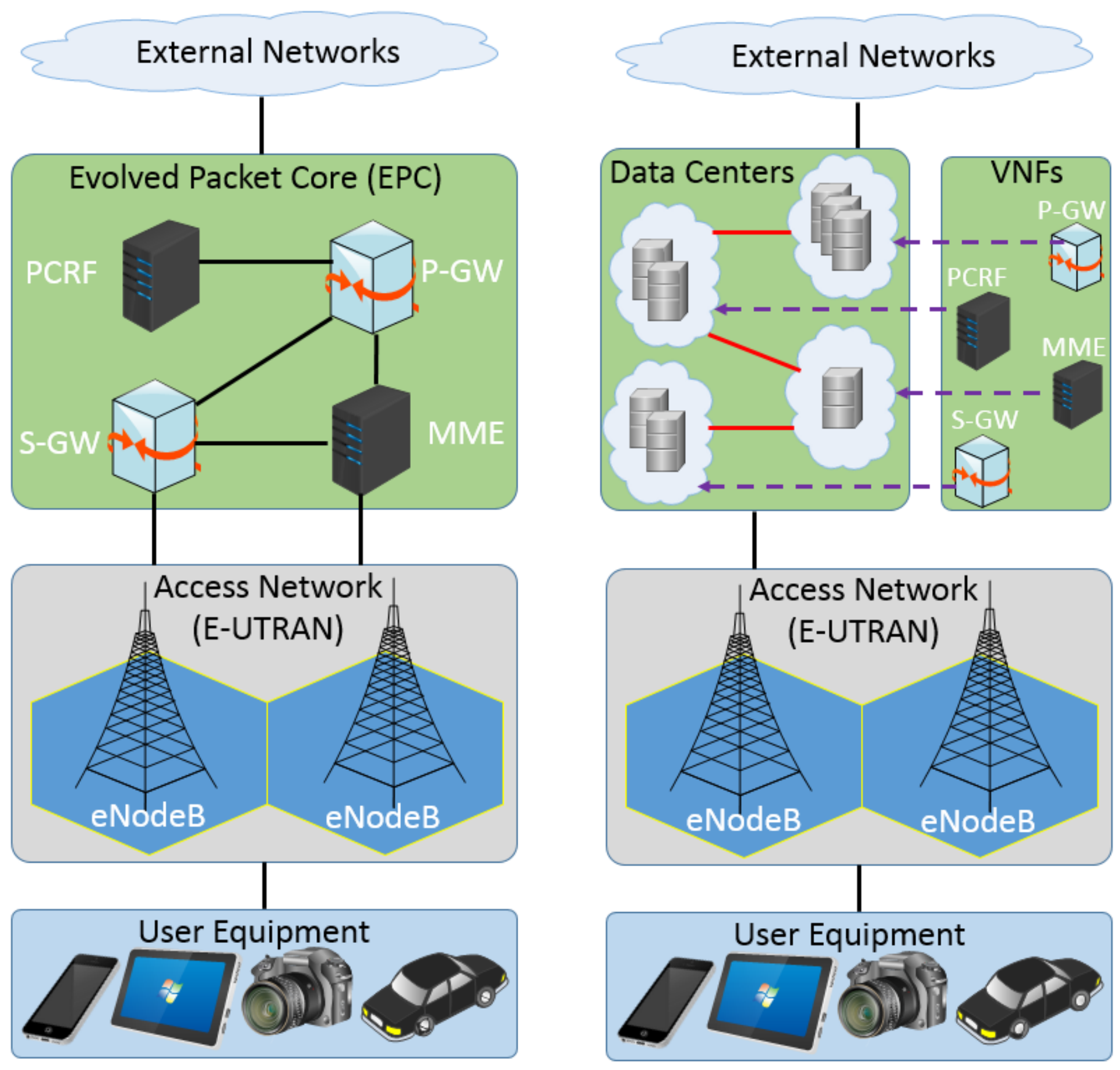}
  \caption{Virtualization of the EPC}
  \label{vepc}
\end{figure}

Virtualizing the \ac{EPC} is another example of \ac{NFV} that has attracted a lot of attention from industry. The \ac{EPC} is the core network for \ac{LTE} as specified by 3GPP \cite{3GPP}. On the left side of Fig. \ref{vepc}, we show a basic architecture of \ac{LTE} without \ac{NFV}. The User Equipment (UE) is connected to the \ac{EPC} over the \ac{LTE} access network (E-UTRAN). The evolved NodeB (eNodeB) is the base station for LTE radio. The EPC performs essential functions including subscriber tracking, mobility management and session management. It is made up of four \acp{NF}: Serving Gateway (S-GW), Packet Data Network (PDN) Gateway (P-GW), Mobility Management Entity (MME), and Policy and Charging Rules Function (PCRF). It is also connected to external networks, which may include the IP Multimedia Core Network Subsystem (IMS). In the current \ac{EPC}, all its functions are based on proprietary equipment. Therefore, even minor changes to a given function may require a replacement of the equipment. The same applies to cases when the capacity of the equipment has to be changed.

On the right side of Fig. \ref{vepc}, we show the same architecture in which the \ac{EPC} is virtualized. In this case, either all functions in the EPC, or only a few of them are transferred to a shared (cloud) infrastructure. Virtualizing the \ac{EPC} could potentially lead to better flexibility and dynamic scaling, and hence allow \acp{TSP} to respond easily and cheaply to changes in market conditions. For example, as represented by the number of servers allocated to each function in Fig. \ref{vepc}, there might be a need to increase user plane resources without affecting the control plane. In this case, \acp{VNF} such as a virtual MME may scale independently according to their specific resource requirements. In the same way, \acp{VNF} dealing with the data plane might require a different number of resources than those dealing with signaling only. This flexibility would lead to more efficient utilization of resources. Finally, it also allows for easier software upgrades on the \ac{EPC} network functions, which would hence allow for faster launch of innovative services.

\subsection{Related Work and Open Questions}
While both industry and academia embrace NFV at unprecedented speeds, the development is still at an early stage, with many open questions. As TSPs and vendors look at the details of implementing NFV and accomplishing its foreseen goals, there are concerns about the realization of some of these goals and whether implementation translates to the benefits initially expected. There are important unexplored research challenges such as testing and validation \cite{PaulVeitch15}, resource management, inter-operability, instantiation, performance of VNFs, etc, that should be addressed. Even areas being explored such as \ac{MANO} still have open questions especially with regard to support for heterogeneity.

There have been recent efforts to introduce NFV, explain its performance requirements, architecture, uses cases and potential approaches to challenges \cite{HanBoGo15}. A discussion of challenges to introducing NFV in mobile networks, with a focus on virtualized evolved packet core is presented in \cite{Hawilo14}, while the reliability challenges of NFV infrastructures are examined in \cite{Cotroneo14}. However, all efforts in current literature are narrow in at least one of the following main ways: (1) with regard to scope, they do not consider important aspects of NFV, such as its relationship with SDN and cloud computing, (2) limited review and analysis of standardization activities, and (3) incomplete descriptions of ongoing research and state-of-the-art efforts and research challenges.

This paper examines the state-of-the-art in NFV and identifies key research areas for future exploration. In addition, we explore the relationship between NFV and two closely related fields, SDN \cite{KreutzAR15} and cloud computing \cite{ZhangCB10}. We also describe the different research and industrial initiatives and projects on NFV, as well as early implementation, proof of concepts and product cases. To the best of our knowledge, this paper presents the most comprehensive state-of-the-art survey on NFV to date.

\subsection{Organization}

The rest of this paper is organized as follows: Section \ref{nfvarch} presents the NFV architecture that has been proposed by ETSI, and discusses its limitations. We propose a reference business model and identify important design considerations in section \ref{bmdc}. In section \ref{related}, we introduce SDN and cloud computing, describing the relationship between them and NFV, as well as current efforts to implement environments involving all of them. In section \ref{soap}, we survey the major projects on NFV as well as early implementations, use cases and commercial products. Based on a qualitative analysis of the state-of-the-art, section \ref{openchallenges} identifies key research areas for further exploration, and section \ref{conclusion} concludes this paper.

\section{\ac{NFV}  Architecture}\label{nfvarch}
\begin{figure}[t]
\centering
\includegraphics[width=8.5cm, height=6.3cm]{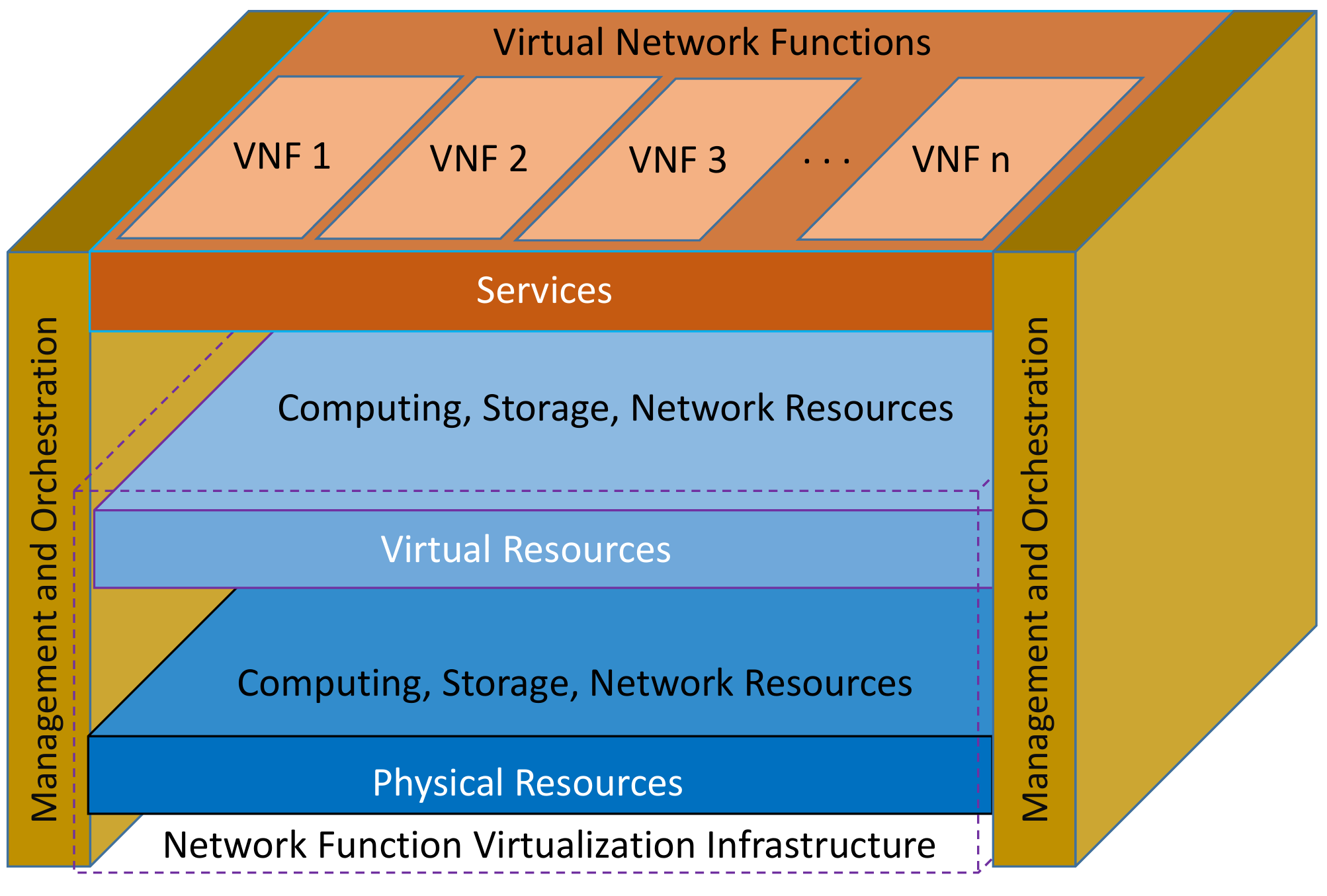}
  \caption{Network Function Virtualization Architecture}
  \label{archit}
\end{figure}

According to ETSI, the NFV Architecture is composed of three key elements: \ac{NFVI}, \acp{VNF} and NFV \ac{MANO} \cite{ETSINFV003}. We represent them graphically in Fig. \ref{archit}. In this section these elements are defined \cite{ETSINFV002, ETSINFV003, NadueauQui14}.

\subsection{NFV Infrastructure (NFVI)} The NFVI is the  combination of both hardware and software resources which make up the environment in which VNFs are deployed. The physical resources include commercial-off-the-shelf (COTS) computing hardware, storage and network (made up of nodes and links) that provide processing, storage and connectivity to VNFs. Virtual resources are abstractions of the computing, storage and network resources. The abstraction is achieved using a virtualization layer (based on a hypervisor), which decouples the virtual resources from the underlying physical resources. In a data center environment, the computing and storage resources may be represented in terms of one or more \acp{VM}, while virtual networks are made up of virtual links and nodes. A virtual node is a software component with either hosting or routing functionality, for example an operating system encapsulated in a VM. A virtual link is a logical interconnection of two virtual nodes, appearing to them as a direct physical link with dynamically changing properties \cite{mijumbi14}. 

\subsection{Virtual Network Functions and Services}  
A \ac{NF} is a functional block within a network infrastructure that has well defined external interfaces and well-defined functional behaviour \cite{ETSINFV003}. Examples of NFs are elements in a home network, e.g. Residential Gateway (RGW); and conventional network functions, e.g. DHCP servers, firewalls, etc. Therefore, a VNF is an implementation of an NF that is deployed on virtual resources such as a VM. A single VNF may be composed of multiple internal components, and hence it could be deployed over multiple VMs, in which case each VM hosts a single component of the VNF \cite{ETSINFV002}. A service is an offering provided by a TSP that is composed of one or more NFs. In the case of NFV, the NFs that make up the service are virtualized and deployed on virtual resources such as a VM. However, in the perspective of the users, the services$-$whether based on functions running dedicated equipment or on VMs$-$should have the same performance. The number, type and ordering of VNFs that make it up are determined by the service's functional and behavioral specification. Therefore, the behaviour of the service is dependent on that of the constituent VNFs. 

\subsection{NFV Management and Orchestration (NFV MANO)} According to the ETSI's MANO framework \cite{ETSINFV009}, NFV MANO provides the functionality required for the provisioning of VNFs, and the related operations, such as the configuration of the VNFs and the infrastructure these functions run on. It includes the orchestration and lifecycle management of physical and/or software resources that support the infrastructure virtualization, and the lifecycle management of VNFs. It also includes databases that are used to store the information and data models which define both deployment as well as lifecycle properties of functions, services, and resources. NFV MANO focuses on all virtualization-specific management tasks necessary in the NFV framework. In addition the framework defines interfaces that can be used for communications between the different components of the NFV MANO, as well as coordination with traditional network management systems such as Operations Support System (OSS) and Business Support Systems (BSS) so as to allow for management of both \acp{VNF} as well as functions running on legacy equipment.\\\\

\noindent \textbf{Discussion:} The ETSI-proposed NFV reference architecture specifies initial functional requirements and outlines the required interfaces. However, the ETSI's scope of work is rather limited, excluding aspects such as control and management of legacy equipment \cite{ETSINFV002}. This could make it difficult to specify the operation and \ac{MANO} of an end-to-end service involving both legacy functions and \acp{VNF}. In addition, standards and/or de-facto best practices and reference implementations of the \acp{VNF}, infrastructure, MANO and detailed definitions of required interfaces are not yet available.

In particular, it can be seen from current NFV solutions that vendors have differing ideas on what constitutes an NFVI and VNFs, and how both of them can be modeled. There remains a number of open questions such as: (1) which \acp{NF} should be deployed in data center nodes, and which ones in operator nodes; (2) which functions should be deployed on dedicated \acp{VM} and which ones in containers\footnote{In fact, even the fact whether containers may be used to host \acp{VNF} and the corresponding ecosystem still needs research \cite{ETSINFV006}.}; (3) what quantity and types of NFVI resources will be required to run specific functions; and (4) operational requirements of environments that involve both \acp{VNF} and those running on legacy equipment.
While many of these questions such as inter-operability and interface definition will be addressed in the second Phase of ETSI's work, time is of the essence. Since both vendors and \acp{TSP} are already investing significantly in NFV, we could reach a point where it is impossible to reverse the vendor-specific solutions.

\section{Business Model and Design Considerations}\label{bmdc}

Using the architecture represented in Fig. \ref{archit}, and based on business models for network virtualization \cite{Chowdhury2010862} and cloud computing \cite{ZhangCB10}, we identify five main players in a \ac{NFV} environment and propose a reference business model that illustrates the possible business relationships between them as shown in Fig. \ref{model}. We also discuss important NFV system design considerations.

\subsection{Business Model}

\subsubsection{Infrastructure Provider (InP)}
InPs deploy and manage physical resources in form of data centers and physical networks. It is on top of these resources that virtual resources may be provisioned and leased through programming interfaces to one or more TSPs. The InPs may also determine how the pool of the available resources are allocated to the TSPs. In NFV, examples of InPs could be public data centers such as those by Amazon, or private servers owned by TSPs. If a given InP is not able to provide resources fully or in part to a given TSPs, negotiations and hence coalitions can be formed with other InPs so as to provision multi-domain VNFs \cite{SamuelJNSA13}.

\subsubsection{Telecommunications Service Provider (TSP)}
\acp{TSP}\footnote{In this paper, we use the term \ac{TSP} to generally mean all service providers. This includes service providers such as Netflix that deploy services with caches in different locations, as well as the traditional TSPs such as Telefonica and Deutsche Telecom.} lease resources from one or more InPs, which they use for running VNFs. They also determine the chaining of these functions to create services for end users. In a more general case, TSPs may sub-lease their virtual resources to other TSPs. In such a case, the reselling \ac{TSP} would take up the role of a InP. In cases where the InP is private or in-house, e.g. provided by TSP network nodes or servers, then the InP and TSP may be one entity.
\begin{figure}[t]
\centering
\includegraphics[width=8.75cm, height=5.5cm]{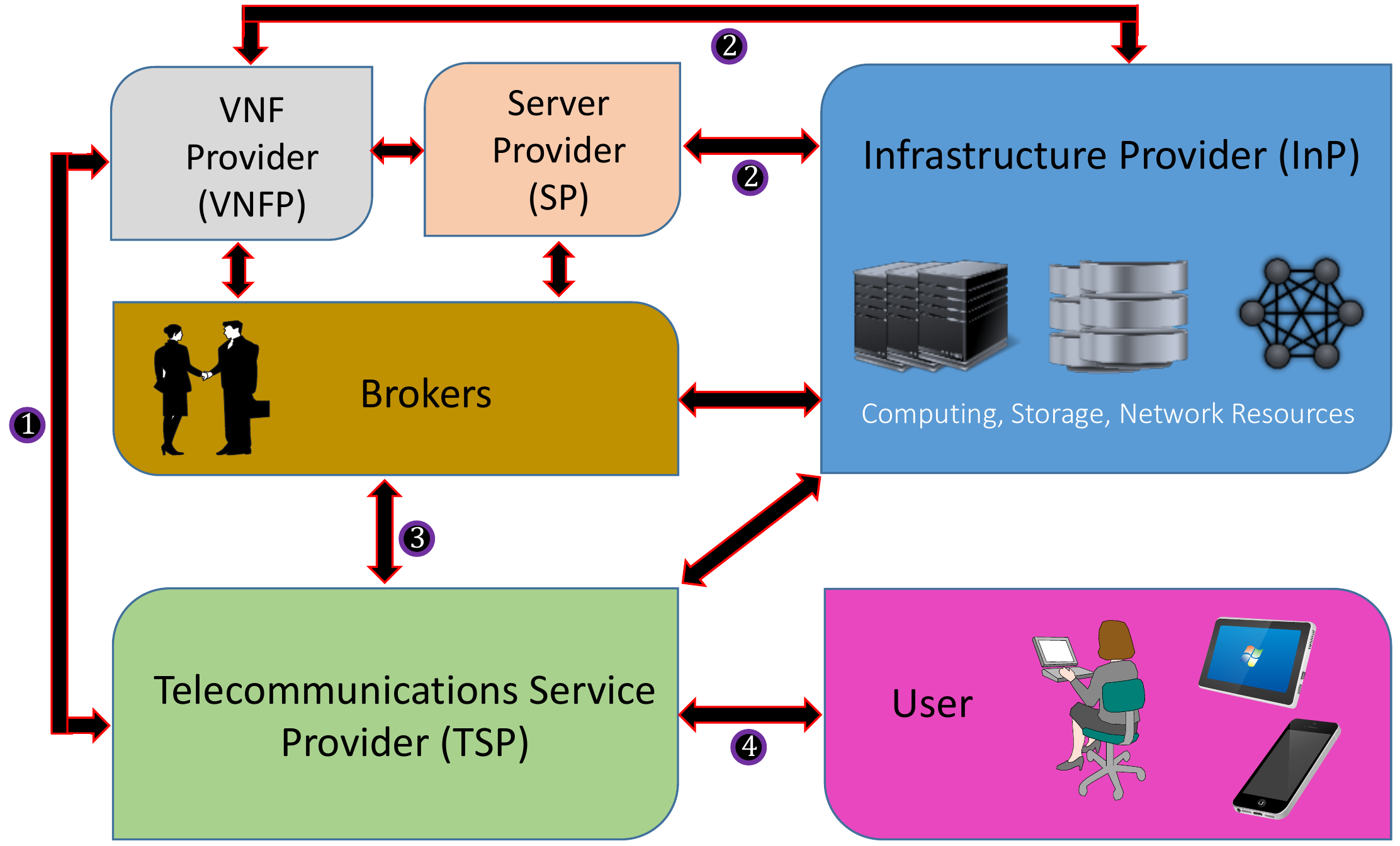}
  \caption{Proposed NFV Business Model}
  \label{model}
\end{figure}

\subsubsection{VNF Providers (VNFPs) and Server Providers (SPs)}
NFV splits the role of traditional network equipment vendors (such as Cisco, Huawei, HP and Alcatel-Lucent) into two: VNFPs and SPs. VNFPs provide software implementations for \acp{NF}. These functions may either be provided directly to \acp{TSP} (via interface 1), or VNFPs could provide them to InPs (via interface 2), who would then provide both infrastructure as well as VNFs to \acp{TSP}. It is also possible that \acp{TSP} develop (some of) their own \acp{NF} (software). In this case, VNFPs and TSPs would be one entity.

In the same way, SPs provide industry standard servers on which VNFs can be deployed. These servers may be provided to InPs (in case the functions will be run in a cloud), or to TSPs (in case the functions will be run in the network nodes of TSPs). It is worth noting that these entities (VNFPs and SPs) may in fact be one company. The main difference is that the functions they provide are not tied to running on equipment with specialized functionality or made by a specific vendor. In other words, a TSP could purchase VNFs from one entity, and servers from a different one.

\subsubsection{Brokers}

In some cases, a \ac{TSP} may need to purchase functions which make up a single service from multiple VNFPs, and/or to deploy and manage the resulting end-to-end services running on resources from multiple InPs. In this case, it may be necessary to have a brokerage role. The brokers would receive resource and/or functions requirements from TSPs and then discover, negotiate and aggregate resources and functions from multiple InPs, VNFPs and SPs to offer them as a service to the TSP. This role is only included in the model for completeness as it may not be required in all cases of the NFV ecosystem.

\subsubsection{End User}
End users are the final consumers of the services provided by TSPs. They are similar to the end users in the existing Internet, except that the existence of multiple services from competing TSPs enables them to choose from a wide range of services. End users may connect to multiple TSPs for different services.

Finally, the arrows in Fig. \ref{model} indicate business relationships or interfaces between the the different entities. For example, VNFPs and/or SPs use interfaces 1 and 2 to negotiate and/or provide VNFs and commodity servers respectively, to TSPs and InPs, while \acp{TSP} use interfaces 3 and 4 for their interactions with brokers and users respectively.

\subsection{NFV Design Considerations}
As NFV matures, it is important to note that it is not only sufficient to deploy NFs over virtualized infrastructures. Network users are generally not concerned with the complexity (or otherwise) of the underlying network. All users require is for the network to allow them access to the applications they need, when they need them. Therefore, NFV will only be an acceptable solution for TSPs if it meets key considerations identified below.

\subsubsection{Network Architecture and Performance}
To be acceptable, NFV architectures should be able to achieve performance similar to that obtained from functions running on dedicated hardware. This requires that all potential bottlenecks at all layers of the stack are evaluated and mitigated. As an example, if VNFs belonging to the same service are placed in different VMs, then there must be a connection between these two VMs, and this connection must provide sustained, aggregated high bandwidth network traffic to the VNFs. To this end, it may be important for the network to be able to take advantage of connections to the network interfaces that are high-bandwidth and low latency due to processor offload techniques such as direct memory access (DMA)\cite{DMA91} for data movement and hardware assist for CRC computation \cite{WWPeter61, Overture15}.

In addition, some \acp{VNF} such as \ac{DPI} are network and compute intensive, and may require some form of hardware acceleration \cite{Byma14} to be provided by the NFVI to still meet their performance goals \cite{Bronstein15}. Some recent efforts \cite{WINDINTEL} have studied the implications of utilizing \acp{DPDK} for running \acp{VNF} and shown that near-native (i.e., similar to non-virtualized) performance for small and large packet processing can be achieved. In addition, Field-Programmable Gate Arrays (FPGAs) have also been shown to enhance performance of \acp{VNF} \cite{Ge26314, Nobach15}. Finally, VNFs should only be allocated the storage and computation resources they need. Otherwise, NFV deployments may end up requiring more resources, and hence there would be no justification for transiting to NFV.

\subsubsection{Security and Resilience}
The dynamic nature of NFV demands that security technologies, policies, processes and practices are embedded in its genetic fabric \cite{ETSINFV011}. In particular, there are two important security risks that should be considered in NFVI designs: (1) functions or services from different subscribers should be protected/isolated from each other. This helps to ensure that functions are resilient to faults and attacks since a failure or security breach in one function/service would not affect another. (2) the NFVI (physical and virtual resources) should be protected from the delivered subscriber services. One way to secure the NFVI is to deploy internal firewalls within the virtual environment \cite{Overture15}. These would allow for the NFV MANO to access to the VNFs without letting malicious traffic from the customer networks into the NFVI. Finally, to make service deployment resilient, it may be necessary for functions that make up the same service not be hosted by physical resources in the same fault or security domain during deployment.

\subsubsection{Reliability and Availability}
Whereas in the IT domain outages lasting seconds are tolerable and a user typically initiates retries, in telecommunications there is an underlying service expectation that outages will be below the recognizable level (i.e. in the order of milliseconds), and service recovery is performed automatically. Furthermore, service impacting outages need to be limited to a certain amount of users (e.g. a certain geography) and network wide outages are not acceptable \cite{ETSINFV010}. These high reliability and availability needs are not only a customer expectation, but often a regulatory requirement, as TSPs are considered to be part of critical national infrastructure, and respective legal obligations for service assurance/business continuity are in place. However, not every function has the same requirements for resiliency: For example, whereas telephony usually has the highest requirements for availability, other services, e.g. Short Messaging Service (SMS), may have lower availability requirements. Thus, multiple availability classes may be defined which should be supported by a NFV framework \cite{ETSINFV010}. Again, functions may be deployed with redundancy to recover from software or hardware failures. 

\subsubsection{Support for Heterogeneity}
The main selling point of NFV is based on breaking the barriers that result from proprietary hardware-based service provision. It is therefore needless to mention that openness and heterogeneity will be at the core of NFV's success. Vendor-specific NFV solutions with vendor-specific hardware and platform capabilities defeat the original NFV concept and purpose. Therefore, any acceptable NFV platform must be an open, shared environment capable of running applications from different vendors. InPs must be free to make their own hardware selection decisions, change hardware vendors, and deal with heterogeneous hardware. In addition, such platforms should be able to shield VNFs from the specifics of the underlying networking technologies (e.g., optical, wireless, sensor etc.) \cite{Alcatel5Issues}. Finally, and equally important, platforms should allow for possibilities of an end-to-end service to be created on top of more than one infrastructural domain without restrictions, and without need for technology specific solutions. While virtualization within a single InP reduces cost, inter-provider NFV enables the ``productization" of the same internal software functions and results in opportunities for revenue growth \cite{ATISNFV15}. As an example, if a mobile user subscribing to given TSP roams into the coverage of another TSP, the user should not be restricted to voice, data and simple messaging services. The real power of NFV would be realized if such a user is able to choose a firewall or security service from the current TSP, or use a combination of functions from the host TSP and others from the one for which he has coverage.

\subsubsection{Legacy Support}
Backward compatibility will always be an issue of high concern for any new technology. NFV is not an exception. 
It is even more important for the telecommunications industry, given that even for a given operator that decides to make the transition to NFV, it may take time for this to be complete, let alone the fact that some operators will do this faster than others. Therefore, support for both physical and virtual NFs is important for operators making the transition to NFV as they may need to manage legacy physical assets alongside virtualized functions for some time. This may necessitate having an orchestration strategy that closes the gap between legacy services and NFV. It is important to maintain a migration path toward NFV, while keeping operators' current network investments in place \cite{Allot2013}. InPs must be able to function in an environment whereby both virtualized and physical network functions operate on the network simultaneously.

\subsubsection{Network Scalability and Automation}
In order to achieve the full benefits of NFV, a scalable and responsive networking solution is necessary. Therefore, while meeting the above design considerations, NFV needs to be acceptably scalable to be able to support millions of subscribers. To give an example, most current NFV proof-of-concepts are based on deploying a VM to host a VNF. Just like a single VM may not be able to meet the requirements of a given function, it is not economical to deploy a VM per NFV, as the resulting VM footprint would be too large, and would lead to scalability problems at the virtualization layer.  However, NFV will only scale if all of the functions can be automated. Therefore, automation of processes is of paramount importance to the success of NFV \cite{Guerzoni12}. In addition, the need for dynamic environments requires that VNFs can be deployed and removed on demand and scaled to match changing traffic. 
\begin{figure*}[t]
\begin{minipage}{.99\textwidth}
\centering
\resizebox{.99\textwidth}{!}
{\includegraphics{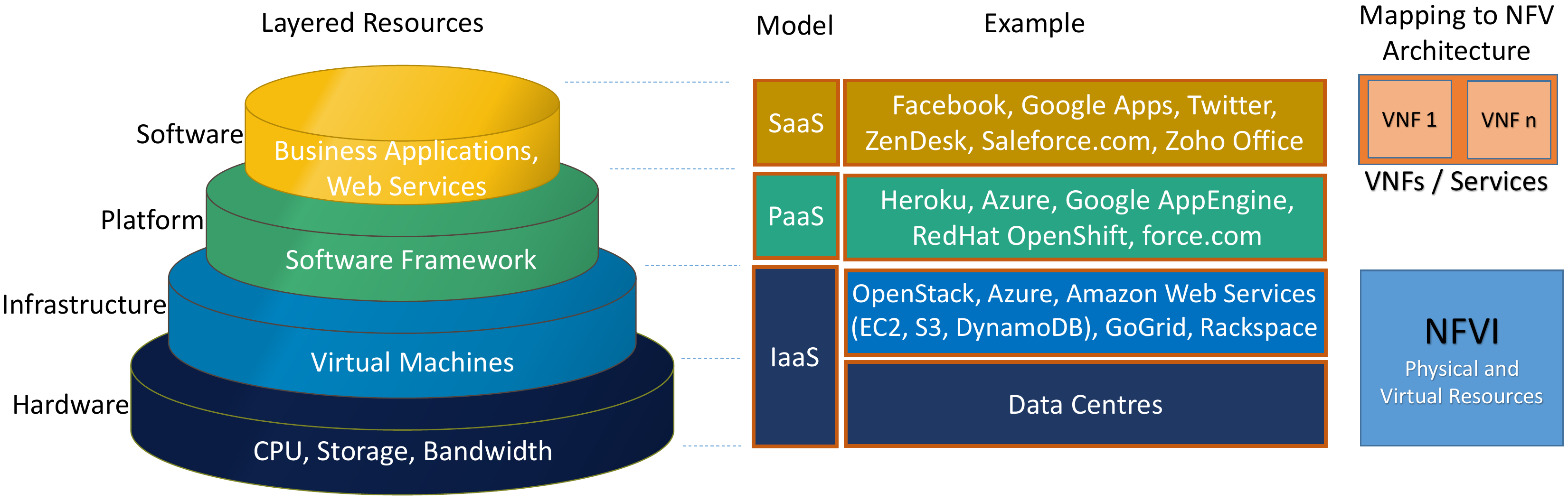}}
  \caption{Cloud Computing Service Models and their Mapping to Part of the NFV Reference Architecture}
  \label{cloudmodels}
\end{minipage}
\end{figure*}

\section{Related Concepts}\label{related}

The need for innovativeness, agility and resource sharing is not new. In the past, the communications industry has invented and deployed new technologies to help them offer new and multiple services in a more agile, cost and resource effective way. In this section, we introduce two such concepts that are closely related to NFV; cloud computing and \ac{SDN}. We also discuss the relationship between NFV and each of them, as well as current attempts to enable all three to work together.

\subsection{Cloud Computing}
According to NIST \cite{NISTCC} cloud computing is ``a model for enabling ubiquitous, convenient, on-demand network access to a shared pool of configurable computing resources (e.g., networks, servers, storage, applications, and services) that can be rapidly provisioned and released with minimal management effort or service provider interaction". In a cloud computing environment, the traditional role of service provider is divided into two: the infrastructure providers who manage cloud platforms and lease resources according to a usage-based pricing model, and  service providers, who rent resources from one or many infrastructure providers to serve the end users \cite{ZhangCB10}. The cloud model is composed of five essential characteristics and three service models \cite{NISTCC}. We briefly introduce these in the following subsections.

\subsubsection{Essential Characteristics of Cloud Computing}
\textit{On-demand self-service}. A consumer can unilaterally provision computing capabilities, such as server time and network storage, as needed automatically without requiring human interaction with each service provider.

\textit{Broad network access}. Capabilities (e.g. compute resources, storage capacity) are available over the network and accessed through standard mechanisms that promote use by heterogeneous thin or thick client platforms (e.g., mobile phones, tablets, laptops, and workstations).

\textit{Resource pooling}. The provider's computing resources are pooled to serve multiple consumers using a multi-tenant model, with different physical and virtual resources dynamically assigned and reassigned according to consumer demand.


\textit{Rapid elasticity}. Capabilities can be elastically provisioned and released, in some cases automatically, to scale rapidly outward and inward commensurate with demand.


\textit{Measured service}. Cloud systems automatically control and optimize resource use by leveraging a metering capability at some level of abstraction appropriate to the type of service (e.g., storage, processing, bandwidth, and active user accounts). 


\begin{table*}[t]
\caption{Comparison of NFV in Telecommunication Networks and Cloud Computing}
\label{SDNCCTable}
\small
\renewcommand{\arraystretch}{1.5}
\centering
\rowcolors{2}{gray!25}{white}
\begin{tabular}{ L{2.2cm} L{8.0cm} L{6.5cm} }
\hline
\bfseries Issue & \bfseries NFV (Telecom Networks) & \bfseries Cloud Computing\\
\hline\hline
Approach  & Service/Function Abstraction & Computing Abstraction\\
Formalization & ETSI NFV Industry Standard Group   & DMTF Cloud Management Working Group \cite{CloudMagt}\\
Latency & Expectations for low latency  & Some latency is acceptable\\
Infrastructure    &  Heterogeneous transport (Optical, Ethernet, Wireless) & Homogeneous transport (Ethernet)\\
Protocol  &  Multiple Control Protocols (e.g OpenFlow \cite{McKeown08}, SNMP \cite{Mauro005}) & OpenFlow \\
Reliability   & Strict 5 NINES availability requirements \cite{5Nines} & Less strict reliability requirements \cite{Vishwanath10}\\
Regulation  &  Strict Requirements e.g NEBS \cite{NEBS} & Still diverse and changing\\
 
\hline
\end{tabular}
\end{table*}

\subsubsection{Cloud Computing Service Models}
The three service models of cloud computing are shown in Fig. \ref{cloudmodels}, and defined below \cite{NISTCC}.

\textit{Software as a Service (SaaS)}. The user is able to use the provider’s applications running on a cloud infrastructure. The applications are accessible from various client devices through either a thin client interface, such as a web browser (e.g., web-based email), or a program interface. 


\textit{Platform as a Service (PaaS)}. The user is able to deploy onto the cloud infrastructure consumer-created or acquired applications created using programming languages, libraries, services, and tools supported by the provider. 


\textit{Infrastructure as a Service (IaaS).} The user is able to provision processing, storage, networks, and other fundamental computing resources where the consumer is able to deploy and run arbitrary software, which can include operating systems and applications.

\subsubsection{Relationship between Cloud Computing and NFV}

In general, \ac{NFV} is not restricted to functions for services in telecommunications. In fact, many IT applications already run on commodity servers in the cloud \cite{Vishwanath10}. However, since most of the promising use cases for NFV originate from the telecommunications industry, and because the performance and reliability requirements of carrier-grade functions are higher than those of IT applications, the discussions in this paper consider that acceptable NFV performance should be carrier-class. In Fig. \ref{cloudmodels}, we have mapped the cloud service models to part of the NFV architecture. It can be observed that IaaS corresponds to both the physical and virtual resources in the NFVI, while the services and VNFs in NFV are similar to the SaaS service model in cloud computing.

Being the cheapest choice for testing and implementation, most NFV proof of concepts and early implementations have been based on deploying functions on dedicated VMs in the cloud. The flexibility of cloud computing, including rapid deployment of new services, ease of scalability, and reduced duplication, make it the best candidate that offers a chance of achieving the efficiency and expense reduction that are motivating TSPs towards NFV. 

However, deploying NFs in the cloud will likely change every aspect of how services and applications are developed and delivered. While work continues to be done with respect to networked clouds and inter-cloud networking \cite{Scharf12, Mandal11}, telecommunication networks differ from the cloud computing environment in at least three ways: (1) data plane workloads in telecom networks imply high pressure on performance, (2) telecom network topologies place tough demands on the network and the need for global network view for management \cite{Lopez14}, (3) the telecom industry requires scalability, five-nines availability and reliability. In traditional telecom networks, these features are provided by the site infrastructure. If NFV should be based on cloud computing, these features need to be replicated by the cloud infrastructure in such a way that they can be orchestrated, as orchestrated features can be exposed through appropriate abstractions, as well as being coupled with advanced support for discoverability and traceability \cite{EricssonCloud14}. It is therefore worth stressing that NFV will require more considerations than just transferring carrier class network functions to the cloud. There is need to adapt cloud environments so as to obtain carrier-class behaviour \cite{Lopez14}. In Table \ref{SDNCCTable}, we summarize the relationship between NFV for telecom networks and cloud computing.
\begin{figure*}[t]
\begin{minipage}{.4\textwidth}
\centering
\resizebox{.80\textwidth}{!}
{\includegraphics{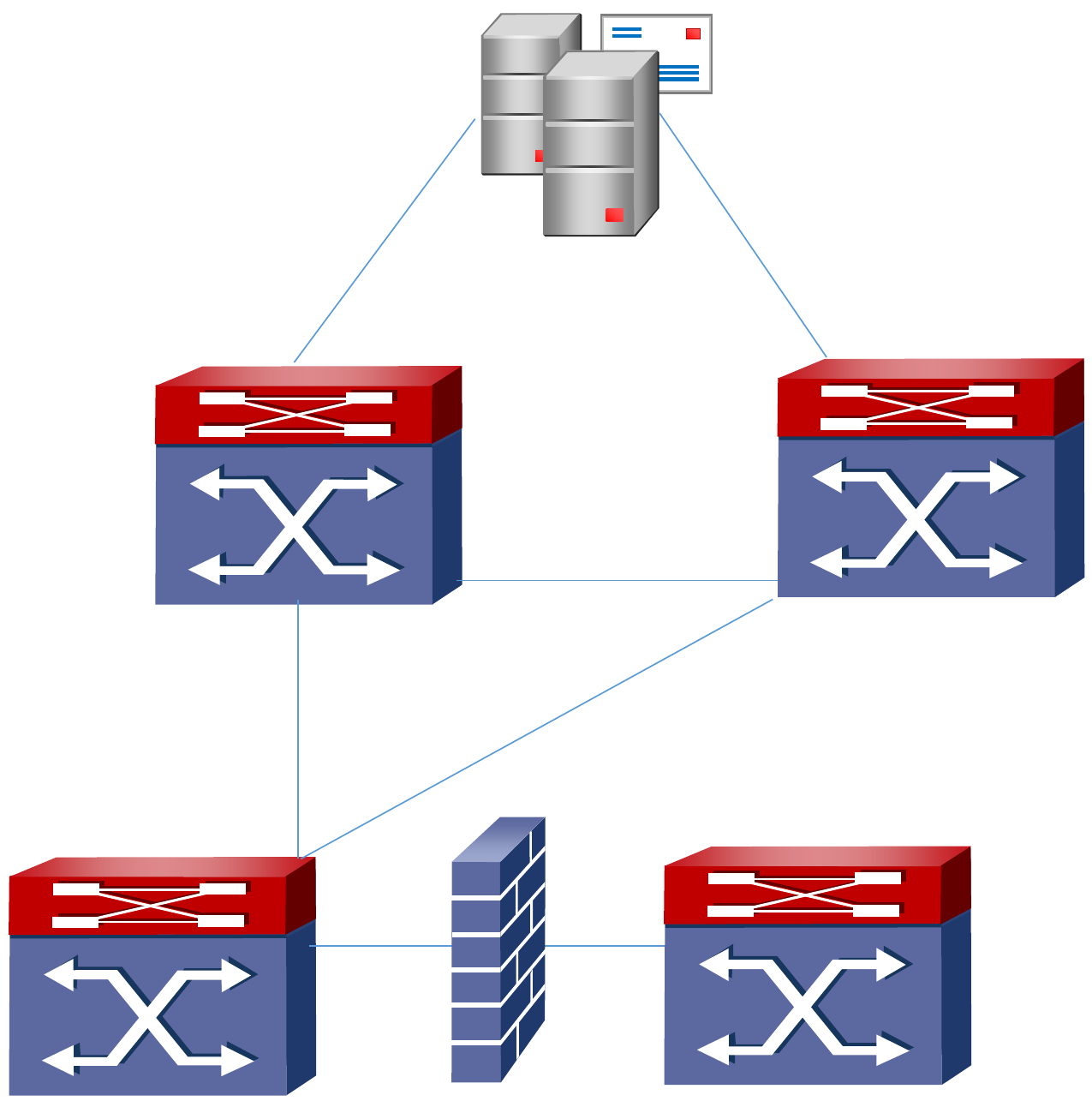}}
  \caption{Distributed Control and Middleboxes (e.g. Firewall, Intrusion Detection, etc.) in Traditional Networks}
  \label{withoutSDNNet}
\end{minipage}
\begin{minipage}{.6\textwidth}
\centering
\resizebox{.99\textwidth}{!}
{\includegraphics{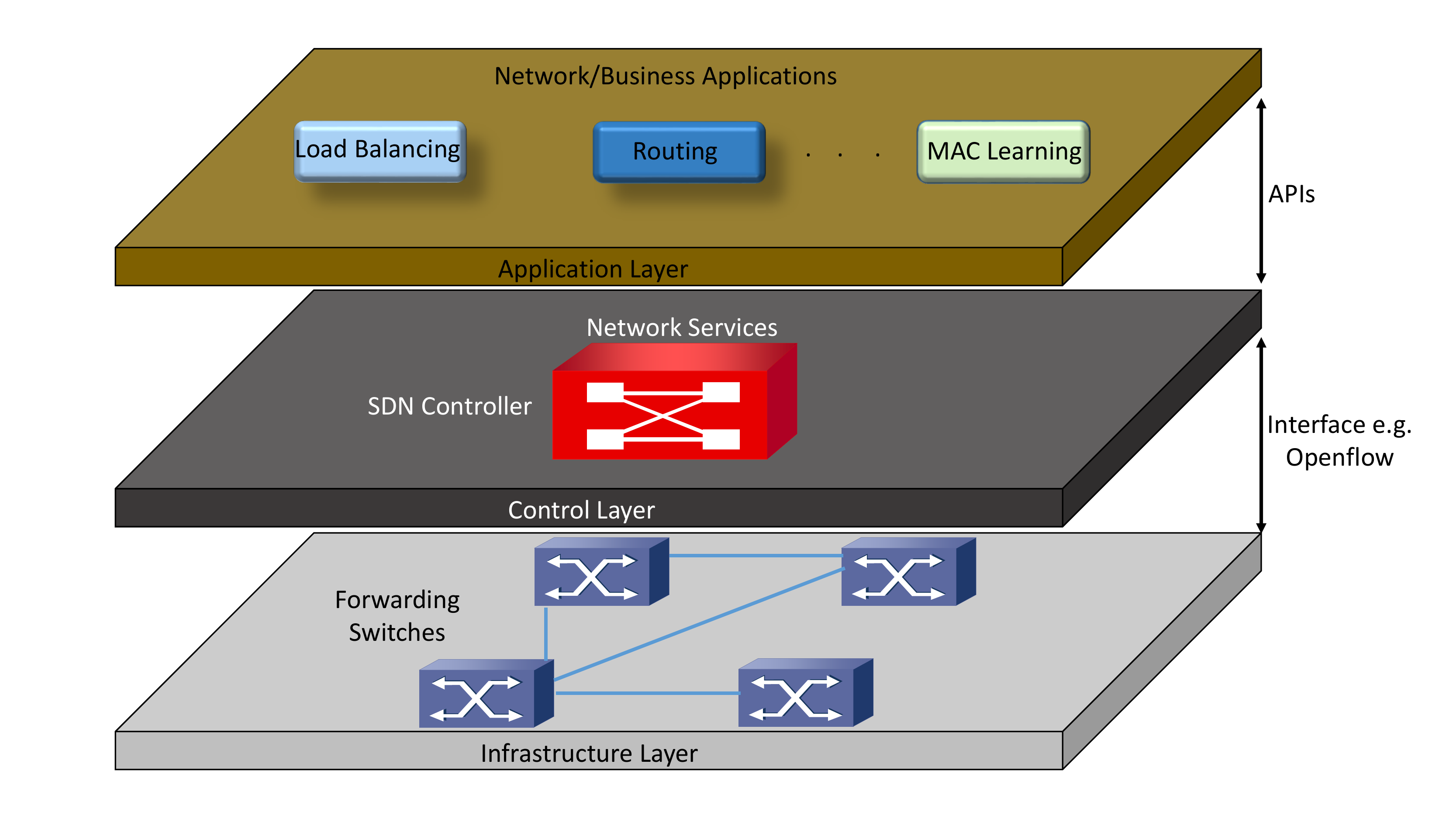}}
  \caption{Logical Layers in a Software Defined Network}
  \label{withSDN}
\end{minipage}
\end{figure*}

\subsubsection{Research on Cloud-based NFV}

In order for NFV to perform acceptably in cloud computing environments, the underlying infrastructure needs to provide a certain number of functionalities which range from scheduling to networking and from orchestration to monitoring capacities.  While OpenStack has been identified as one of the main components of a cloud-based NFV architectural framework, it currently does not meet some NFV requirements. For example, through a gap analysis in \cite{ETSIOPEN14}, it was noted that, among other gaps, OpenStack neither provides detailed description of network resources including \ac{QoS} requirements, nor supports a resource reservation service and consequently it does not provide any interface for resource reservation. 

In addition, through measurements some performance degradation has been reported \cite{Callegati14}. Some efforts have already been dedicated to study the requirements needed to make the performance of cloud carrier-grade \cite{GlithoR14, Antti14, Jamjoom14}. In particular, OpenANFV \cite{Ge26314} proposes an OpenStack-based framework which uses hardware acceleration to enhance the performance of \acp{VNF}. The author's efforts are motivated by the observation that for some functions (e.g., DPI, network deduplication (Dedup) and \ac{NAT}), industry standard servers may not achieve the required levels of performance. Therefore, OpenANFV aims at providing elastic, automated provisioning for hardware acceleration to \acp{VNF} in OpenStack. To this end, the tested \acp{VNF} (DPI, Dedup and NAT) were allowed access to a predefined set of accelerated behavior and to communicate through a hardware-independent interface with the hypervisor to configure the accelerator. The authors reported performances 20, 8 and 10 times better for DPI, Dedup and NAT respectively.

\subsection{Software Defined Networking (SDN)}

SDN \cite{FeiHuQi14} is currently attracting significant attention from both academia and industry as an important architecture for the management of large scale complex networks, which may require re-policing or re-configurations from time to time. As shown in Figures \ref{withoutSDNNet} and \ref{withSDN}, SDN decouples the network control and forwarding functions. This allows network control to become directly programmable via an open interface (e.g., ForCES \cite{Doria10forwardingand}, OpenFlow \cite{AKLara14}, etc) and the underlying infrastructure to become simple packet forwarding
devices (the data plane) that can be programmed.

While the SDN control plane can be implemented as pure software which runs on industry-standard hardware, the forwarding plane requires an SDN agent \cite{SDNARCH14}, and may therefore require to be implemented in specialized hardware. However, depending on the performance and capacity needs of the SDN networking element, and depending on whether specialized hardware transport interfaces are required, the forwarding plane may also be implemented on commodity servers \cite{MetaswictchMT14}. For example, VMware's NSX platform \cite{VMWARENSX} includes a virtual switch (vSwitch) and controller both of which implement SDN protocols without requiring specialized hardware.

SDN has the potential to dramatically simplify network management and enable innovation and evolution \cite{BAANunes14}. According to the \ac{ONF} \cite{SDNONF}, SDN addresses the fact that the static architecture of conventional networks is ill-suited for the dynamic computing and storage needs of today's data centers, campuses, and carrier environments. The SDN architecture is \cite{Jarschel14}:

\textit{Programmable.} SDN makes network control directly programmable since control is decoupled
from forwarding functions. This programmability can be used to automate network configuration in such a way that network administrators can run `SDN apps' that help to optimize particular services such as VoIP so as to ensure a high \ac{QoE} for phone calls.


\textit{Agile.} Abstracting control from forwarding lets administrators dynamically adjust network-wide
traffic flow to meet changing needs. This makes the network more agile since logic is now implemented in a software running on commodity hardware, which has shorter release cycles than device firmware.

\textit{Centrally managed.} Network intelligence is (logically) centralized in software-based SDN
controllers that maintain a global view of the network, which appears to applications and policy
engines as a single, logical switch.

\textit{Open standards-based and vendor-neutral.} When implemented through open standards,
SDN simplifies network design and operation because instructions are provided by SDN
controllers instead of multiple, vendor-specific devices and protocols.

\begin{table*}[htb]
\caption{Comparison of Software Defined Networking and Network Function Virtualization Concepts}
\label{SDNNFVTable}
\small
\renewcommand{\arraystretch}{1.5}
\centering
\rowcolors{2}{gray!25}{white}
\begin{tabular}{ L{3.0cm} L{6.9cm} L{6.9cm} } \hline
\bfseries  Issue & \bfseries NFV (Telecom Networks) & \bfseries Software Defined Networking\\ \hline \hline
Approach  & Service/Function Abstraction & Networking Abstraction\\
Formalization & ETSI   & \ac{ONF}\\
Advantage  & Promises to bring flexibility and cost reduction& Promises to bring unified programmable control and open interfaces\\
Protocol  & Multiple control protocols (e.g SNMP, NETCONF)& OpenFlow is de-facto standard\\
Applications run  & Commodity servers and switches & Commodity servers for control plane and possibility for specialized hardware for data plane \\
Leaders  & Mainly Telecom service providers & Mainly networking software and hardware vendors\\
Business Initiator & Telecom service providers & Born on the campus, matured in the data center\\
\hline
\end{tabular}

\end{table*}


\subsubsection{Relationship between SDN and NFV} NFV and SDN have a lot in common since they both advocate for a passage towards open software and standard network hardware. Specifically, in the same way that NFV aims at running \acp{NF} on industry standard hardware, the SDN control plane can be implemented as pure software running on industry standard hardware. In addition, both NFV and SDN seek to leverage automation and virtualization to achieve their respective goals. In fact, NFV and SDN may be highly complimentary, and hence combining them in one networking solution may lead to greater value. For example, if it is able to run on a VM, an SDN controller may be implemented as part of a service chain. This means that the centralized control and management applications (such as load balancing, monitoring and traffic analysis) used in SDN can be realized, in part, as \acp{VNF}, and hence benefit from NFV's reliability and elasticity features. In the same way, SDN can accelerate NFV deployment by offering a flexible and automated way of chaining functions, provisioning and configuration of network connectivity and bandwidth, automation of operations, security and policy control \cite{Cui14}. It is however worth stressing that most of the advantages expected from both NFV and SDN are promises that have not been proven yet.

However, SDN and NFV are different concepts, aimed at addressing different aspects of a software-driven networking solution. NFV aims at decoupling \acp{NF} from specialized hardware elements while SDN focuses on separating the handling of packets and connections from overall network control. As stated by the ONF in the description of the SDN architecture \cite{SDNARCH14}, ``the NFV concept differs from the virtualization concept as used in the SDN architecture. In the SDN architecture, virtualization is the allocation of abstract resources to particular clients or applications; in NFV, the goal is to abstract \acp{NF} away from dedicated hardware, for example to allow them to be hosted on server platforms in cloud data centers".  It can be observed that the highest efforts in promoting and standardizing SDN is in data center and cloud computing areas while telecom carriers are driving similar efforts for NFV. Finally, an important distinction is that while NFV can work on existing networks because it resides on servers and interacts with specific traffic sent to them, SDN requires a new network construct where the data and control planes are separate. We summarize the relationship between SDN and NFV in Table \ref{SDNNFVTable}.

\subsubsection{Research on SDN-based NFV}

There is currently a lot of work involving the combination of SDN and NFV to enhance either of them; including: a ForCES-based framework \cite{haleplidis2014sdnnfv}, NFV-based monitoring for SDN \cite{Choi2014}, an abstraction model for both the forwarding model and for the network functions \cite{haleplidis2014sdnnfv}. As these efforts show, the unique demands of NFV will potentially necessitate a massively complex forwarding plane, blending virtual and physical appliances with extensive control and application software, some of it proprietary \cite{ONF14Feb}. There are two major aspects of SDN that may need to be improved in order to meet the requirements of NFV: the Southbound API (mainly OpenFlow), and controller designs. We discuss advances in each of these two aspects below.


\paragraph{Southbound API}

OpenFlow is the de-facto implementation of a southbound API for SDN. However, before we consider NFV support, even in current SDN environments OpenFlow is by no means a mature solution \cite{FiaWYen14}. Since OpenFlow targets L2-L4 flow handling, it has no application-layer protocol support and switch-oriented flow control. Therefore, users have to arrange additional mechanism for upper-layer flow control. Furthermore, executing a lot of flow matching on a single switch (or virtual switch) can cause difficulties in network tracing and overall performance degradation \cite{Kawashima12}.

Therefore, OpenFlow will have to be extended to include layers L5-L7 to be able to support NFV. Basta et al.   \cite{Kellerrer13} investigated the current OpenFlow implementation in terms of the basic core operations such as \ac{QoS}, data classification, tunneling and charging, concluding that there is a need for an enhanced OpenFlow to be able to support some functions in an NFV environment. In an implementation of a virtual EPC function \cite{ETSIUseCases}, \cite{KempfJ12} extends OpenFlow 1.2 by defining virtual ports to allow encapsulation and to allow flow routing using the GTP Tunnel Endpoint Identifier (TEID). 

Finally, while OpenFlow assumes a logically centralized controller, which ideally can be physically distributed, most current deployments rely on a single controller. This does not scale well and can adversely impact reliability. In addition, network devices in an NFVI require collaboration to be able to provide services, which cannot currently be provided by SDN. There is therefore still a need to improve SDN by considering distributed architectures \cite{Tootoonchian106, YaziciVSu12}. It may also be important for TSPs, InPs and ETSI to consider other possible solutions such as NETCONF \cite{YuJALj10}.

\paragraph{Controller Design}

While there are multiple controllers that may be used in an SDN environment, all of them require improvements to be able to support NFV requirements, especially with regard to distributed network management and scalability. OpenNF \cite{Gember14, Kawashima12} proposes a control plane that allows packet processing to be redistributed across a collection of NF instances, and provides a communication path between each NF and the controller for configuration and decision making. It uses a combination of events and forwarding updates to address race conditions, bound overhead, and accommodate a variety of NFs. \cite{Batalle13} also designed a protocol to implement the communication between the controllers and the VNFs. Finally, \cite{MallaSam15} proposes an architecture that considers the control of both SDN and NFV.

OpenDaylight \cite{opendaylight} is one of the few SDN control platforms that supports a broader integration of technologies in a single control platform \cite{KreutzAR15}. A collaborative project hosted by the Linux Foundation, OpenDaylight is a community-led and industry-supported open source framework to accelerate adoption, foster new innovation and create a more open and transparent approach to SDN and NFV. The objective of the OpenDaylight initiative is to create a reference framework for programmability and control through an open source SDN and NFV solution. The argument of OpenDaylight is that building upon an open source SDN and NFV controller enables users to reduce operational complexity, extend the life of their existing infrastructure hardware and enable new services and capabilities only available with SDN. 


\subsection{Summary: NFV, SDN and Cloud Computing}

\begin{figure}[t]
\centering
\includegraphics[width=9.15cm, height=8.0cm]{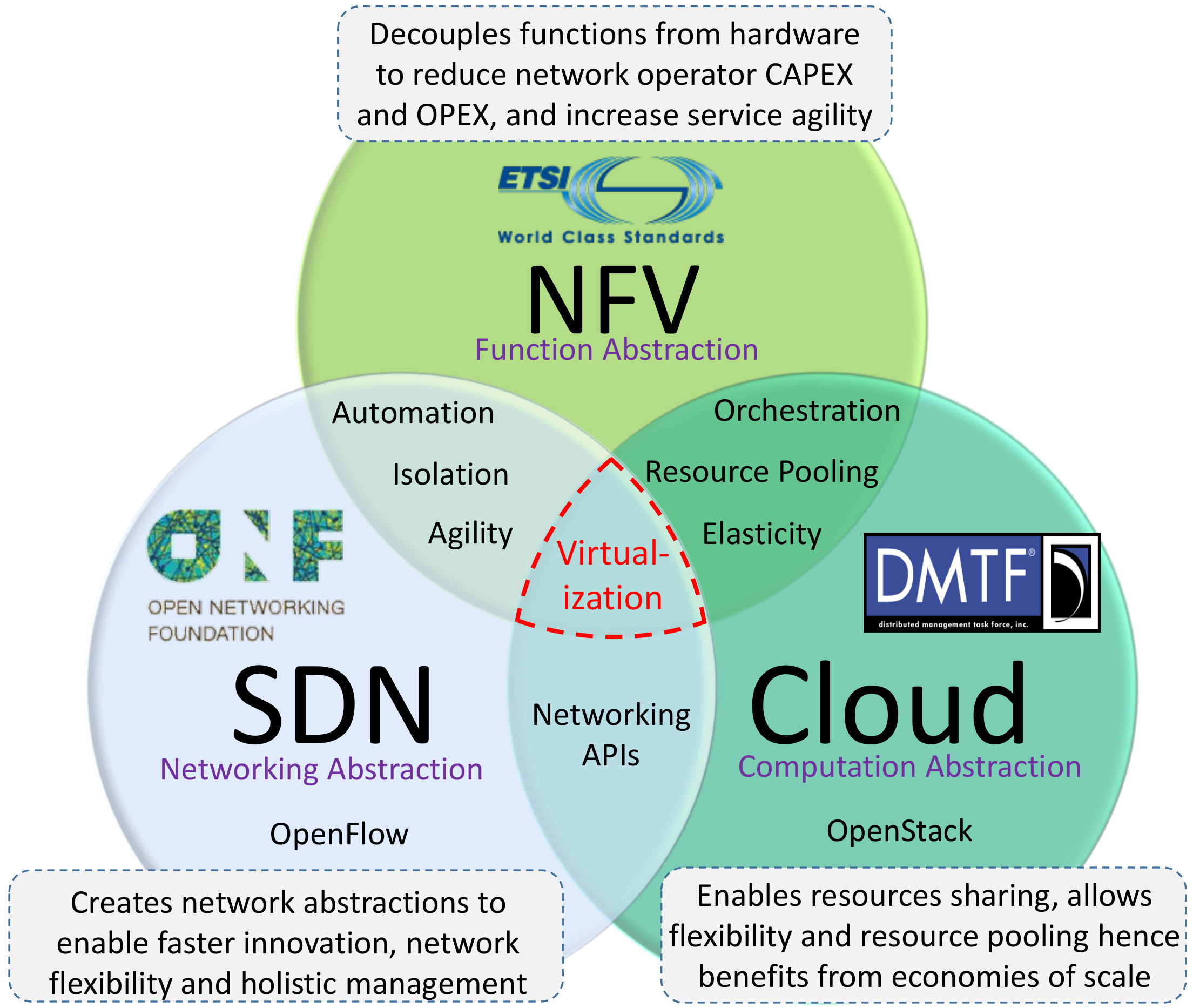}
  \caption{Relationship between NFV, SDN \& Cloud Computing}
  \label{nscareas}
\end{figure}

To summarize the relationship between NFV, SDN, and cloud computing, we use Fig. \ref{nscareas}\footnote{It is worth remarking that OpenFlow is not the only SDN protocol. In the same way, OpenStack is not the only cloud computing platform. The reason we present only these two in Fig. \ref{nscareas} is that, as already mentioned, they have received more attention in general, and with regard to NFV.}. We observe that each of these fields is an abstraction of different resources: compute for cloud computing, network for SDN, and functions for NFV. The advantages that accrue from each of them are similar; agility, cost reduction, dynamism, automation, resource scaling etc. 

The question is not whether NFs will be migrated to the cloud, as this is in fact the general idea of NFV. It is whether the cloud will be a public one like Amazon, or if TSPs will prefer to user private ones distributed across their infrastructure. Either way, work will have to be done to make the cloud carrier-grade in terms of performance, reliability, security, communication between functions, etc.

On the other hand, NFV goals can be achieved using non-SDN mechanisms, and relying on the techniques currently in use in many data centers. However, approaches relying on the separation of the control and data forwarding planes as proposed by SDN can enhance performance, simplify compatibility with existing deployments, and facilitate operation and maintenance procedures. In the same way, NFV is able to support SDN by providing the infrastructure upon which the SDN software can be run. Finally, the modern variant of a data center (the cloud and it's self-service aspect) relies on automated management that may be obtained from SDN and NFV. In particular, aspects such as network as a service, load balancing, firewall, VPN etc. all run in software instantiated via APIs 

\section{State-of-the-art}\label{soap}

As the ETSI continues work on NFV, several other standards organizations, academic and industrial research projects and vendors are working in parallel with diverse objectives, and some of them in close collaboration with the ETSI. In this section, we explore these NFV activities.

\subsection{NFV Standardization Activities}

\subsubsection{IETF Service Function Chaining Working Group}
Functions in a given service have strict chaining and/or ordering requirements that must be considered when decisions to place them in the cloud are made. The Internet Engineering Task Force (IETF) \cite{IETF15} has created the Service Function Chaining Working Group (IETF SFC WG) \cite{IETFSFC} to work on function chaining. The IETF SFC WG is aimed at producing an architecture for service function chaining that includes the necessary protocols or protocol extensions to convey the service function chain (SFC) and service function path information \cite{IETFSFCDOCS} to nodes that are involved in the implementation of service functions  and SFCs, as well as mechanisms for steering traffic through service functions.
\subsubsection{IRTF NFV Research Group (NFVRG)}
The Internet Research Task Force (IRTF) has created a research group, NFVRG \cite{IRTFNFVRG}, to promote research on NFV. The group is aimed at organizing meetings and workshops at premier conferences and inviting special issues in well-known publications. The group focuses on research problems associated with NFV-related topics and on bringing a research community together that can jointly address them, concentrating on problems that relate not just to networking but also to computing and storage aspects in such environments. 


\subsubsection{ATIS NFV Forum}
The ATIS NFV Forum \cite{ATISNFV15} is an industry group created by the Alliance for Telecommunications Industry Solutions (ATIS), a North American telecom standards group. The group is aimed at developing specifications for NFV, focusing on aspects of NFV which include inter-carrier inter-operability and new service descriptions and automated processes. ATIS NFV Forum plans to develop technical requirements, the catalog of needed capabilities and the service chaining necessary for a third party service provider or enterprise to integrate the functions into a business application. This process is expected to result in creation of specifications that are complementary with existing industry work products and that extend the current environment for inter-provider NFV. The forum also engages open source activities for the implementation of these capabilities in software.

\subsubsection{Broadband Forum}
The Broadband Forum (BB Forum) \cite{BBForum} is an industry consortium dedicated to developing broadband network specifications. Members include telecommunications networking and service provider companies, broadband device and equipment vendors, consultants and independent testing labs (ITLs). BB Forum collaborates with the ETSI after agreeing a formal liaison relationship in 2013. The BB Forum is working on how NFV can be used in the implementation of the multi-service broadband network (MSBN). To this end, the forum has many work items in progress, including: migrating to NFV in the context of TR-178 (WT-345), introducing NFV into the MSBN (SD-340), virtual business gateway (WT-328), flexible service chaining (SD-326) \cite{BBForumWTs}.
%

\begin{table*}[t]
\caption{Summary of Network Function Virtualization Standardization Efforts}
\label{standardssummary}
\small
\renewcommand{\arraystretch}{1.5}
\centering
\rowcolors{2}{gray!25}{white}

\begin{tabular}{ C{1.8cm}  C{3.9cm}  C{1.8cm}  L{9.2cm}   } \hline
 & Description & Focus Area & Description of NFV-Related Work \\ \hline\hline

ETSI & Industry-led ETSI Standards Group  & NFV & NFV architectural framework, infrastructure description, \ac{MANO}, security and trust, resilience and service quality metrics.\\

3GPP SA5 & 3GPP's Telecom Management working group  & Mobile Broadband & Working in liaison with the ETSI. Studying the management of virtualized 3GPP network functions.\\

IETF SFC WG & IETF Working Group &   NFV & To propose a new approach to service delivery and operation, an architecture for service function chaining, management and security implications.\\

IRTF NFVRG & IRTF Research Group   &  NFV & Organizing NFV-related research activities in both academia and industry through workshops, research group meetings etc. at premier conferences.\\

ATIS NFV Forum & Industry-led Standards Group   & NFV & Developing specifications for NFV, focusing on inter-carrier interoperability. \\

ONF & Industry-led consortium for standardization of OpenFlow   & SDN & Standardizing the OpenFlow protocol and related technologies. Defines OpenFlow as the first standard communications interface defined between the control and forwarding layers of an SDN architecture.\\

DMTF OVF & Industry-led consortium  & Cloud & DMTF's OVF and the CIM may be used as one option for capturing some or all of the VNF package and/or VDU \cite{ETSINFV009} Descriptor.\\

BB Forum & Industry-led consortium that develops broadband network specifications   & NFV in Broadband Networks & Collaborating with the ETSI to achieve a consistent approach and common architecture for the infrastructure needed to support VNFs.\\ \hline

\end{tabular}
\end{table*}
\subsubsection{Standardization of Related Paradigms}\label{dtmfovfref}

In addition to the NFV standardization efforts, other bodies continue to work on standardization of related fields, SDN and cloud computing, which may also play a significant role in the success of NFV. The DMTF defined the Open Virtualization Format (OVF) \cite{DMTFOVF} to address the portability and deployment of physical machines, virtual machines and appliances. OVF enables the packaging and secure distribution of virtual machines or appliances, providing cross-platform portability and simplified deployment across multiple platforms including cloud environments. OVF has adopted by both ANSI as a National Standard and ISO as the first international virtualization and cloud standard. It takes advantage of the DMTF's Common Information Model (CIM) \cite{DMTFCIM}, where appropriate, to allow management software to clearly understand and easily map resource properties by using an open standard.  OVF and CIM may be used as one option for capturing some or all of the VNF package and/or Virtual Deployment Unit (VDU) descriptor \cite{ETSINFV009, HuaweiSDO}. Although OVF does a great job enabling the provisioning of workloads across various clouds, it is still insufficient for new era cloud applications and runtime management.

In the same way, the ONF is standardizing the OpenFlow protocol and related technologies. ONF defines OpenFlow as the first standard communications interface defined between the control and forwarding layers of an SDN architecture. ONF has more than 123 member companies, including equipment vendors, semiconductor companies, computer companies, software companies, telecom service providers, etc. 

In Table \ref{standardssummary}, we summarize all the activities in the standardization of NFV and related technologies. 
In general, it can be said that there is sufficient involvement of standards bodies in NFV activities. While many of them work in liaison with the ETSI, some of them such as ATIS and 3GPP SA5 have identified and are working on specific aspects of NFV that have not yet been sufficiently developed by the ETSI. What remains to be seen is whether the output in terms of standards will match with the speed at which vendors and TSPs propose NFV solutions.

\subsection{Collaborative NFV Projects}

\subsubsection{Zoom}

Zero-time Orchestration, Operations and Management (ZOOM) \cite{TMFZoom} is a TM Forum project aimed at defining an operations environment necessary to enable the delivery and management of \acp{VNF}, and  identifying new security approaches that will protect NFVI and \acp{VNF}. To achieve these objectives, the project regularly conducts a range of hands-on technology demos each of which is developed from what they call a catalyst project. Each catalyst project is sponsored by one or more network operators and equipment and software vendors in a real-world demo. The project currently runs about 9 catalysts with a focus on NFV aspects such as  end-to-end automated management, security orchestration, function and service modeling, and using big data technologies and open software principles for workload placement.

\subsubsection{Open Platform for NFV (OPNFV)}
OPNFV \cite{OPNFV} is an open source project founded and hosted by the Linux Foundation, and composed of TSPs and vendors. It aims to establish a carrier-grade, integrated, open source reference platform to advance the evolution of NFV and to ensure consistency, performance and inter-operability among multiple open source components. The first outcome of the project is referred to as OPNFV Arno \cite{OPNFVArno}, and was released in June 2015. The release provides an initial build of the \ac{NFVI} and Virtual Infrastructure Manager (VIM) components of the ETSI architecture. It is developer-focused, and can therefore be used to explore NFV deployments, develop \ac{VNF} applications, or to evaluate NFV performance and for use case-based testing. In particular, Arno has capabilities for integration, deployment and testing of components from other projects such as Ceph, KVM, OpenDaylight, OpenStack and Open vSwitch. In addition, end users and developers can deploy their own or third party VNFs on Arno to test its functionality and performance in various traffic scenarios and use cases.

\subsubsection{OpenMANO}
OpenMANO \cite{OpenMANO} is an open source project led by Telefonica, which is aimed at implementing ETSI's NFV MANO framework.
Specifically, it attempts to address aspects related to performance and portability by applying Enhanced Platform
Awareness (EPA) \cite{EPA} principles. The OpenMANO architecture is made up of three main components: openmano, openvim and a graphical user interface (GUI). OpenMANO has a northbound interface (openmano API), based on REST, where MANO services are offered including the creation and deletion of VNF templates, VNF instances, network service templates and network service instances. Openvim is a lightweight, NFV-specific virtual infrastructure manager implementation directly interfacing with the compute and storage nodes in the NFVI, and with an openflow controller in order to create the infrastructural network topology. It offers a REST-based northbound interface (openvim API) where enhanced cloud services are offered including the lifecycle management of images, flavors, instances and networks. The REST interface of openvim is an extended version of the OpenStack API to accommodate EPA.

\subsubsection{Mobile Cloud Networking (MCN)} MCN \cite{eumcn15} is a consortium consisting of network operators, cloud providers, vendors, university and research institutes, as well as SMEs. The objective is to cloudify all components of a mobile network operation such as: the access - \ac{RAN}; the core - \ac{EPC}; the services - IP Multimedia Subsystem (IMS), Content Delivery Networks (CDN) and Digital Signage (DSS); the Operational Support Systems (OSS) and the Business Support Systems (BSS).

\subsubsection{UNIFY} UNIFY \cite{JohnCsad13} is aimed at researching, developing and evaluating the means to orchestrate, verify and observe end-to-end service delivery from home and enterprise networks through aggregation and core networks to data centers. To this end, the project plans to develop an automated, dynamic service creation platform, leveraging a fine-granular service chaining architecture. They will also create a service abstraction model and a service creation language to enable dynamic and automatic placement of networking, computing and storage components across the infrastructure. Finally, they will develop a global orchestrator with optimization algorithms to ensure optimal placement of elementary service components across the infrastructure. 

\subsubsection{T-NOVA} T-NOVA  \cite{t-nova} aims at promoting the NFV concept, by proposing an enabling framework, allowing operators not only to deploy VNFs for their own needs, but also to offer them to their customers, as value added services. For this purpose, T-NOVA leverages SDN and cloud management architectures to design and implement a management/orchestration platform for the automated provision, configuration, monitoring and optimization of Network Functions-as-a-Service (NFaaS) over virtualized network/IT infrastructures.

\subsubsection{CONTENT}  CONTENT \cite{CONTENT} is an EU funded project aimed at offering a network architecture and overall infrastructure solution to facilitate the deployment of conventional cloud computing as well as mobile cloud computing. The main objectives of the project include: (1) proposing a cross-domain and technology virtualization solution allowing the creation and operation of infrastructure slices including subsets of the network and computational physical resources, and (2) supporting dynamic end-to-end service provisioning across the network segments, offering variable \ac{QoS} guarantees, throughout the integrated network.\\
\begin{table*}[!htpb]
\caption{Summary of Network Function Virtualization Projects}
\label{projectsummary}
\small
\renewcommand{\arraystretch}{1.4}
\centering
\rowcolors{2}{gray!25}{white}
\begin{tabular}{ L{1.8cm}  C{2.2cm}  C{2.0cm}   C{2.1cm}  L{7.8cm} } \hline
 & Project  Type  &  Leader and/or Funding &  Focus Areas  & Main Objective \\ \hline \hline
ZOOM &      Association of SPs   &   TM Forum  &  NFV   & Enable more rapid deployment of services by automating the provisioning process and modernizing OSS/BSS models.\\
OPNFV &    Collaborative Project  &   Linux Foundation  &    NFV  & Build an open source reference platform to advance the evolution of NFV. \\

OpenMANO & Vendor Project & Telefonica & SDN, NFV &Implementation of ETSI's \ac{MANO} framework. \\

MCN &       Research Project  &   European Union &   SDN, NFV   & Cloudify all components of a mobile network operation.\\
UNIFY &  Research Project &    European Union &  NFV   & Develop an automated, dynamic service creation platform, leveraging fine-granular service chaining.    \\
T-NOVA &    Research Project &    European Union &   SDN, NFV & Design and implement a MANO platform for NFV.    \\

CONTENT &    Research Project &    European Union &   Mobile Networks, Cloud  & Providing a technology platform interconnecting geographically distributed computational resources that can support a variety of Cloud and mobile Cloud services.    \\
OpenStack &  Working Group &    OpenStack Foundation &   Cloud, NFV  & Identify requirements needed to deploy, manage, and run telecom services on top of OpenStack. \\
OpenDaylight    &   Collaborative Project  &     Linux Foundation & SDN, NFV  & Develop an open platform for SDN and NFV.  \\ \hline

\end{tabular}
\end{table*}

\noindent \textbf{Summary:}
To summarize, in Table \ref{projectsummary} we present all the projects giving their main objective, their focus with respect to NFV and related areas, and entities leading or funding them. All these projects are guided by the proposals coming out of the standardization described earlier, in particular ETSI, 3GPP and DMTF. It is interesting to observe that all the three industrial projects (ZOOM, OPNFV and OpenMANO) surveyed are focused on \ac{MANO}. This underlines the importance of \ac{MANO} in \ac{NFV}. MANO is a critical aspect towards ensuring the correct operation of the NFVI as well as the VNFs. Just like the decoupled functions, NFV demands a shift from network management models that are device-driven to those that are aware of the orchestration needs of networks which do not only contain legacy equipment, but also VNFs. The enhanced models should have improved operations, administration, maintenance and provisioning focused on the creation and lifecycle management of both physical and virtualized functions. For NFV to be successful, all probable MANO challenges should be addressed at the current initial specification, definition and design phase, rather than later when real large scale deployments commence.

\subsection{NFV Implementations}\label{soa}

In order to demonstrate the possibility to implement the ideas proposed by NFV, and to determine performance characteristics, a number of use cases for NFV, mostly based on those defined by ETSI \cite{ETSIUseCases}, have already been implemented. These have mainly been based on implementing single virtual functions such as routing \cite{JBattellle14}, Broadband Remote Access Server \cite{Masutani14}, policy server \cite{Bulan14}, deep packet inspection \cite{Bremler14}, \ac{EPC} \cite{Gebert2014, MallaSam15}, \ac{RAN} \cite{MugenPeng14, RuiWang14, ChihLin14, Sabella13}, monitoring \cite{Choi2014}, CPE \cite{Vilalta14, Vilalta15, vCPEArinet, vCPEAnuta, vCPEcalsoftlabs, Aleksic14}, GPRS \cite{NagyMart14} and access control \cite{jacobdeploying}, in cloud environments. All these originate from the research community. Perhaps not surprisingly, the biggest implementations have arisen from equipment vendors. In the remainder of this section, we introduce some key NFV implementations and products from industry.




\subsubsection{HP OpenNFV}
The HP OpenNFV \cite{HPOPENNFV} is a platform, based on HP's NFV Reference Architecture, upon which services and networks can be dynamically built. The HP NFV Reference Architecture is aligned towards providing solutions to each of the functional blocks defined in the ETSI architecture, as a starting point. The NFVI and VNFs parts of the architecture mainly include HP servers and virtualization products, while MANO is based on three solutions; NFV Director, NFV Manager, and Helion OpenStack. The NFV Director is an orchestrator that automatically manages the end-to-end service, by managing its constituent VNFs. It also performs global resource management, allocating resources from an appropriate pool based on global resource management policies. VNF managers are responsible for the VNFs lifecycle actions, for example, by deciding to scale in or out. It also includes a Helion OpenStack cloud platform for running VNFs.
\subsubsection{Huawei NFV Open Lab}
The Huawei NFV Open Lab \cite{Huawei} is aimed at providing an environment to ensure that NFV solutions and carrier grade infrastructure are compatible with emerging NFV standards and with the OPNFV \cite{OPNFV}. The lab is dedicated to being open and collaborative, expanding joint service innovations with partners, and developing the open eco-system of NFV to aggregate values and help customers achieve business success. They also plan to collaborate with the open source community to innovate on NFV technologies to provide use cases for multi-vendors inter-operability around NFVI, and VNF-based services.

\subsubsection{Intel Open Network Platform (Intel ONP)}

Intel ONP \cite{IntelOPN} is an ecosystem made up of several initiatives to advance open solutions for \ac{NFV} and \ac{SDN}. The initiatives are focused on Intel product development (such as the Intel ONP Server), participation in open source development and standardization activities and collaborations with industry for proof of concepts and trials.

The main result of the ONP so far is the Intel ONP Server. This is a reference architecture that integrates open-source and hardware ingredients optimized for SDN/NFV. It is aimed at enabling manageability by exposing health, state, and resource availability, for optimal workload placement and configuration. Its software stack consists of released open-source software based on the work done in community projects, including contributions provided by Intel. Some of the key open-source software ingredients forming the Intel ONP Server software stack are OpenStack, OpenDaylight, DPDK, Open vSwitch, and Linux KVM.

\subsubsection{CloudNFV} 

CloudNFV \cite{CloudNFV} is an NFV, SDN and cloud computing platform resulting from cooperation between six companies (6WIND, CIMI Corporation, Dell, EnterpriseWeb, Overture Networks, and Qosmos). CloudNFV proposed their own NFV architecture \cite{CloudNFV} which is made up of 3 main elements: active virtualization, NFV orchestrator, and NFV Manager. Active virtualization is a data model which represents all aspects of services, functions and resources. The VNF orchestrator has policy rules, which, combined with service orders and the status of available resources, determines the location of the functions that make up the service as well as connections between them. The VNF Manager uses a data/resource model structured according to TMF rules and the concept of ``derived operations" is used to manage VNFs. Derived operations are used to integrate the status of available resources with the resource commitments for functions of a given NFV service. The main difference between the ETSI NFV MANO and CloudNFV is that unlike the former, the latter considers both management and orchestration as applications that can run off a unified data model.

\subsubsection{Alcatel-Lucent CloudBand}
Alcatel-Lucent's CloudBand \cite{CloudBand} is a two-level platform implementing NFV. First, it includes nodes that provide resources like VMs and storage, and then, the CloudBand Management System which is the functional heart of the process. It operates as a work distributor that makes hosting and connection decisions based on policy, acting through cloud management APIs. Virtual functions are deployed using \textit{recipes} that define \textit{packages} of deployable components and instructions for their connection. The recipes can be used to set policies and determine how specific components are instantiated and then connected. The platform uses the Nuage SDN technology \cite{NuageNet15} and its related links to create an agile connection framework for the collection of nodes and functions, and to facilitate traffic management.

Alcatel-Lucent recently teamed with RedHat \cite{CloudBand34} such that the latter could fill the gaps required to use CloudBand and OpenStack to promote the inclusion of more NFV requirements in the OpenStack upstream and hence build a solution that is optimized for telco NFV environments. Within this collaboration, the CloudBand node uses the RedHat Enterprise Linux OpenStack platform as the VIM.

\subsubsection{Broadcom Open NFV}
The Broadcom Open NFV platform \cite{Broadcom} is aimed at accelerating creation of NFV applications across multiple system on chip (SoC) \cite{KhanAA04} processors, and to allow system vendors to be able to migrate virtual functions between platforms based on various vendor solutions. Broadcom's platform supports open API standards such as Linaro's Open Data Plane (ODP)\cite{ODP15} to access acceleration components for scaling critical functionality and reducing time-to-market. The ETSI has recently accepted a VNF state migration and inter-operability proof of concept in which Broadcom is demonstrating an implementation of an EPC and migrating the virtual function state from operating on one instruction set architecture (ISA) to a different ISA.

\subsubsection{Cisco Open Network Strategy}

Cisco's Open Network Strategy (OPN) \cite{CiscoESP} includes an Evolved Services Platform (ESP) and an Evolved Programmable Network (EPN). The ESP and EPN include a service orchestrator, a VNF manager, and a SDN controller, all of which are aimed at providing implementations for some of the functional blocks of ETSI's MANO framework. The service orchestrator is responsible for providing the overall lifecycle management at the network service level. The VNF manager provides scalable, automated VNF lifecycle management, including the creation, provisioning, and monitoring of both Cisco and third-party VNFs. The VNF manager is also responsible for the scale-up and scale-down of the VNFs based on dynamic and fluctuating service demands.
It uses cloud-computing resource managers such as OpenStack and VMware at the VIM layer to configure and provision
compute and storage resources across multi-vendor data center networks. Finally, the SDN Controller is responsible for connecting the virtualized services (a VNF or a set of chained VNFs) to the service provider VPNs, the Internet, or both. It is designed around open standards and APIs and uses a holistic systems-based approach to manage multi-vendor and multi-tenant
data centers, and a common policy-based operating model to reduce costs.

%
%
%
%
%
%
\begin{table*}[t]
\caption{Summary of State-of-the-art NFV Implementations}
\label{SOASummary}
\small
\renewcommand{\arraystretch}{1.4}
\centering
\rowcolors{2}{gray!25}{white}

\begin{tabular}{ C{1.8cm}  L{8.5cm}  C{3.6cm}  C{2.6cm} C{2.6cm} } \hline
 &  Functionality & Platform &  Driving Standards  \\ \hline\hline


HP OpenNFV & Open standards-based NFV reference architecture, labs as a sandbox in which carriers and equipment vendors can test  vEPC.  & OpenStack &  ETSI  \\

NFV Open Lab & Supports the development of NFV infrastructure, platforms and services. &  OpenStack, OpenDaylight &  ETSI \\

Intel ONP & Provides developers with a validated template for quickly developing and showcasing next-generation, cloud-aware network solutions.  & OpenStack, OpenDaylight & 3GPP or TMF     \\

CloudNFV & Provides a platform for virtual network service creation, deployment, and management.  & OpenStack & TMF and ETSI     \\

Alcatel CloudBand & Can be used for standard IT needs as well as for CSPs who are moving mobile networks into the cloud.  & Red Hat Linux OpenStack Platform & ETSI    \\

BroadBand NFV & Migrate virtual functions between platforms based on various vendor solutions.  &  & ETSI    \\

Cisco ONS &  Automated service delivery, improved network and data center use, fast deployment of personalized offerings. & OpenStack, OpenDaylight & ETSI       \\

F5 SDAS &  Extensible, context-aware, multi-tenant system for service provisioning & OpenStack, BIG-IP, BIG-IQ \cite{f5} &   IETF, 3GPP, GSMA, ETSI, ONF     \\

ClearWater &  SIP-based call control for voice and video communications and for SIP-based messaging applications. & Apache Cassandra, Memcached & 3GPP IMS, ETSI TS     \\

Overture vSE &  Host multiple VNFs in one box, Accelerate service creation,
activation and assurance, Decrease inventory and management costs, Optimize service flexibility, Eliminate trucks rolls & Linux Overture Ensemble OSA \cite{Overture15}, OpenStack &       \\ \hline

\end{tabular}
\end{table*}

\subsubsection{F5 Software Defined Application Services}

F5 Software Defined Application Services (F5 SDAS) \cite{f5, FrankYue14, f5SDAS} provides Layer 4-7 capabilities to supplement existing Layer 2-3 network and compute initiatives such as SDN. It enables service injection, consumption, automation, and orchestration across a unified operating framework of pooled resources. It is is comprised of three key components: (1) The application service platform supports programmability of both control and data paths. It is extensible and enables new service creation. (2) The application services fabric provides core services such as scalability, service isolation, multi-tenancy, and integration with the network, and (3) Application services, which are the heart of F5 SDAS, are a rich catalog of services across the application delivery spectrum.

\subsubsection{ClearWater} 
ClearWater \cite{ClearWater} is an open source implementation of an IMS built using web development methods to provide voice, video and messaging services to users. It leans heavily on established design patterns for building and deploying scalable web applications, adapting these design patterns to fit the constraints of SIP and IMS. In particular, all components scale out horizontally using simple, stateless load-balancing. In addition, long-lived state is not stored on cluster nodes, avoiding the need for complex data replication schemes. Instead, long-lived state is stored in back-end service nodes using cloud-optimized storage technologies such as Cassandra. Finally, interfaces between the front-end SIP components and the back-end services use RESTful web services APIs.
Interfaces between the various components use connection pooling with statistical recycling of connections to ensure load is spread evenly as nodes are added and removed from each layer.

Metaswitch \cite{Metaswitch} contributed the initial code base for the ClearWater project to software developers and systems integrators, and continues to drive the evolution of the code base.

\subsubsection{Overture Virtual Service Edge (vSE)}

Overture vSE \cite{Overture15} is an open carrier Ethernet platform for hosting VNFs at the service edge. It allows TSPs to instantly deploy on-demand VNFs at the customer premise. It combines carrier Ethernet access with the benefits of virtualization, openness and software-defined services. The result is a single platform for both services and network access, which allows for VNFs to be turned up, down, expanded and removed dynamically so that compute and storage resources are used only when needed. Additionally, it supports multiple wireline and wireless connections to the WAN, allowing access to all end customer locations.

The platform implements an Ethernet access as a \ac{VNF}, and is based on a virtualization platform comprising a Linux KVM/QEMU hypervisor, an optimized virtual switch, and includes supports for OpenStack integration with another product - the Ensemble Service Orchestrator.\\

\noindent \textbf{Summary:} In Table \ref{SOASummary} we summarize the different state-of-art implementations stating their functionality, the standards bodies they closely follow and platforms on which they run. It is worth remarking that although NFV is gaining momentum, it is still an emerging technology and solutions based on final specifications, and widespread deployments for end-users may take a few years to appear. As the survey above shows, many organizations are investing in and are willing to test NFV-based solutions. In addition, it can be observed from these early implementations and platforms, that two aspects re-appear in a big number of them: (1) the high focus on open source, and (2) the ability of current SDN and cloud technologies to support NFV.

%
%
%

\section{Research Challenges}\label{openchallenges}

Even with all the anticipated benefits, and despite the immense speed at which it is being accepted by both academia and industry, NFV is still in early stages. There still remain important aspects that should be investigated and standard practices which should be established. This section discusses crucial research directions that will be invaluable as NFV matures.

\subsection{Management and Orchestration}
The deployment of NFV will greatly challenge current management systems and will require significant changes to the way networks are deployed, operated and managed. Such changes are required, not just to provide network and service solutions as before, but also to exploit the dynamism and flexibility made possible by NFV \cite{Keeney14, Bondan14}. It will likely lead to scenarios where functions that provide a service to a given customer are scattered across different server pools. The challenge then will be to have an acceptable level of orchestration to make sure that on a per service (or user) level, all the required functions are instantiated in a coherent and on-demand basis, and to ensure that the solution remains manageable \cite{Bronstein14}.

ETSI is working on a \ac{MANO} framework \cite{ETSINFV009} required for the provisioning of VNFs, and the related operations, such as the configuration of the VNFs and the infrastructure these functions run on. In a related effort, Cloud4NFV \cite{Soares14, JoaoSoares15} has proposed an end-to-end management platform for VNFs, which is based on the ETSI architectural specification. Clayman et al. \cite{Clayman14} describe an architecture based on an orchestrator that ensures the automatic placement of the virtual nodes and the allocation of network services on them supported by a monitoring system that collects and reports on the behaviour of the resources. NetFATE \cite{riccobenenetwork} proposes an orchestration approach for virtualized functions, taking into account the service chains needed by traffic flows and the desired \ac{QoE}. In addition, other \ac{MANO} frameworks and architectures have been proposed in \cite{Mainiel14, Shen14, WenyuShen15, Donaldo14, Bolla14, Giotis15}.

However, there are still some open issues. Current approaches are focused on NFV management, without considering the management challenges in SDN \cite{SDNMGT15}. While traditional management approaches must be improved to accommodate each one of them, the demands for management are even higher in environments including both. In such cases. we no longer just need to create dynamic traffic flows, but the switching points (locations of functions) are also changing dynamically. Therefore, a complete management solution should combine requirements from both SDN and NFV.
\begin{table*}[!ht]
\caption{Summary of Changes in Energy Consumption from Virtualizing Network Functions}
\label{gwatttable}
\small
\renewcommand{\arraystretch}{1.4}
\centering
\rowcolors{2}{gray!25}{white}

\begin{tabular}{ L{2.9cm}  C{3.2cm}  C{2.2cm}  C{2.0cm} C{2.2cm} C{3.2cm}} \hline
 &  Traffic (EXABYTES/MONTH) & Total Efficiency (MBITS/J) &  Total Power (MWATTS) & Power Savings (MWATTS) & Cummulative Savings (2013 - 2018) GJ  \\ \hline\hline
\textbf{Baseline Network}& $1,153.05$&$0.0328510$&$116,203$ & $0.0$ &  \\
Virtual EPC& $1,153.05$& $0.0422222$& $92,159.8$&$24,044.1$ & $5.0\times 10^9$ \\
Virtual CPE& $1,205.11$ &$0.0352130$ &$113.500$ &$2,703.63$ & $5.5\times 10^9$  \\
Virtual RAN& $1,227.88$&$0.0463708$ &$89,599.5$ &$26,604.4$ &$7.5\times 10^9$  \\
Virtual Video CDN& $810.22$& $0.0346562$& $80,029.3$& $36,174.6$& $7.5\times 10^9$ \\
Virtual Broadband Network Gateway& $1,169.69$& $0.0333016$& $116.260$& $-76.794$& $-1.7\times 10^7$ \\
Virtual Provider Edge& $1,151.91$& $0.0328255$& $116,180$& $22.9517$& $3.8\times 10^6$ \\ \hline
\end{tabular}
\end{table*}

In addition, support for inter-operability is a key requirement for NFV. However, looking at the ETSI MANO framework, most effort has been on defining intra-operator interfaces, without clear guidelines on inter-operability. This is why, while current vendor products are ``based on the ETSI MANO framework", most of them use  custom models and/or representation for functions and services. Furthermore, the need for dynamism in function means that functions will likely be moved from one VM to another. This underscores the importance of a higher focus on possibilities of an availability monitoring mechanism as part of the end-to-end management solution. Finally, while the ETSI-proposed NFV MANO framework considers the management and orchestration requirements of both virtualized and non-virtualized functions via interfaces to traditional network management functions OSS/BSS, the relationship between them is yet to be fully defined \cite{HeavyReading}

\subsection{Energy Efficiency}
Since energy bills represent more than 10\% of \acp{TSP}' \ac{OPEX} \cite{GWATT}, reduced energy consumption is one of the strong selling points of NFV. The argument is that with the flexibility and ability to scale resource allocations up and down, as traffic demands ebb and flow, \acp{TSP} could potentially reduce the number of physical devices operating at any point, and hence reduce their energy bills. Yet, NFV will likely make data centers an integral part of telecommunication networks. According to an analysis in the SMARTer 2020 report from GeSI \cite{GESI}, the cloud, if it were a country, would rank 6th in the world in terms of its energy demand, and yet this demand is expected to increase by 63\% by 2020 \cite{CLICKCLEAN}. While some progress on energy efficient cloud computing has been made, the fast growing energy needs of data centers continue to receive a lot of attention \cite{USDCS, BeloglazovBLZ11}. Therefore, there is an urgent need to study whether NFV will meet its energy savings expectations, or whether$-$like the NFs$-$the energy consumption will just be transferred to the cloud.

China Mobile recently published \cite{ChihLin14} their experiences in deploying a \ac{C-RAN}. One of the tests was performed on their 2G and 3G networks, where it was observed that by centralizing the RAN, power consumption could be reduced by 41\% due to shared air-conditioning. In addition, Shehab et al. \cite{Shehab13} analyzed the technical potential for energy savings associated with shifting U.S. business software to the cloud. The results suggested a substantial potential for energy savings. In fact, the authors noted that if all U.S. business users shifted their email, productivity software, and CRM software to the cloud, the primary energy footprint of these software applications could be reduced by as much as 87\%.

In order to determine the possible effect of energy consumption on the evolution to VNFs, Bell Labs has recently extended its G.W.A.T.T. tool \cite{GWATT}. The tool is able to show the effect of virtualizing different network functions based on forecasts for traffic growth. G.W.A.T.T. divides the network into six domains (Home \& Enterprise, Access \& Aggregation, Metro, Edge, Core and Service Core \& Data Centers). Each network domain can be edited to select different network models and technologies and hence analyze its energy impact. Based on the tool's default settings and using EPC network models for 2015, the tool shows that total network energy efficiency is $0.0422222$ MBITS/J, total energy consumption is $92,159.8$ MWATTS, and that the energy savings resulting from virtualizing the EPC would be $24,044.1$ MWATTS. For the same use case, the tool showed that the total energy savings over a five year period (using 2013 as baseline) would be $5.0\times 10^9$ GJ, and that the energy efficiency of the core network $1.86393$ MBITS/J. The results for some other NFV use cases, including those for the baseline network\footnote{A baseline network is one where all functions are run in physical equipment, using the tool's default technologies and settings.} are summarized in Table \ref{gwatttable}. However, while the tool is an important step in attaching numbers to the energy savings expected from NFV, it can still be improved. In particular, it does not yet have a detailed technical documentation. For example, Cisco's visual networking index \cite{Cisco15} forecasts that annual global IP traffic will reach 1000 exabytes in 2016. Based on this, the (monthly, 2015) traffic values in Table \ref{gwatttable} seem to be too high, yet it is currently not possible to know how these values are derived. 

Therefore, we expect that the energy efficiency of cloud based NFs will continue to receive attention. NFV will put InPs under even more pressure to manage energy consumption \cite{Bolla14} not to only to cut down energy expenses, but also to meet regulatory and environmental standards. Topics with regard to energy efficient hardware which could allow reductions in CPU speeds and partially turning off some hardware components, more energy-aware function placement, scheduling and chaining algorithms, will be important. An example could be to track the cheapest prices for energy costs and adapt the network topology and/or operating parameters to minimize the cost of running the network \cite{Cui14}. However, all these should be carefully considered to ensure that there is a balance in the trade-off between energy efficiency and function performance or service level agreements.

\subsection{NFV Performance}

The concept of \ac{NFV} is to run \acp{NF} on industry standard servers. This means that server providers should produce equipment without knowledge of the characteristics of functions that could run on them in future. In the same way, VNF providers should ensure that the functions will be able to run on commodity server. This raises the question of whether functions run on industry standard servers would achieve a performance comparable to those running on specialized hardware, and whether these functions would be portable between the servers \cite{Cui14}. Finding answers to these questions has been another focus of the ETSI, and resulted into a ``Performance \& Portability Best Practises" specification \cite{ETSINFVperf14}. The specification gives performance test results on NFV use cases such as DPI, C-RAN, BRAS, etc. The results proved that if ``best practices were followed" it was not only possible to achieve high performance (upto 80 Gbps for a server) in a fully virtualized environment, but that the performance was predictable, consistent and in vendor-agnostic manner, leveraging features commonly available in current state-of-the-art servers \cite{Cui14}. 

In a related effort, results from China Mobile's \ac{C-RAN} deployment \cite{ChihLin14} indicated that the Common Public Radio Interface (CPRI) \cite{CPRI13} over a wavelength-division multiplexing (WDM) front-haul transport solution gives ideal performance, with no impact on radio performance. The tests also verified the feasibility of using a general purpose platform (GPP) and the NFV implementation. In particular, a GPP based C-RAN prototype with the ability to support as many as 90 TD-LTE carriers, 15 FDD-LTE carriers and 72 GSM carriers was developed. The prototype demonstrated a similar level of performance to the traditional DSP/FPGA based systems.

However, performance at high speeds is an issue even in non-virtualized \acp{NF} \cite{Nobach15, Napotech14}. Therefore, techniques such as hardware acceleration will also be important for NFV. In fact, hardware acceleration has been shown to improve the performance of some \acp{VNF}. Ge et. al \cite{Ge26314} determine that for some functions (e.g. DPI, Dedup and NAT), industry standard servers may not achieve the required levels of performance. From the authors' tests, a virtualized Dedup could only achieve 267 Mbps throughput in each core at most. It was also proved by Yamazaki et. al \cite{Koji14} who reported achieving a better performance and energy efficiency by deploying a virtualized DPI on Application Specific Instruction-set Processor (ASIP) rather than commodity servers.

Therefore, there are some high performance \acp{NF} that may be difficult to virtualize without degradation in performance. While hardware acceleration may be used for such functions, such specialization is against the concept of \ac{NFV} which aims at high flexibility. There should be defined ways of managing the trade-off between performance and flexibility. It will also be appropriate to have phased migrations to NFV where those functions that have acceptable performance are virtualized first and allowed to run alongside unvirtualized or physical ones.

\subsection{Resource Allocation}
To achieve the economies of scale expected from NFV, physical resources should be used efficiently. It has been shown that default deployment of some current use cases may result in sub-optimal resource allocation and consumption \cite{PaulVeitch15}.

This calls for efficient algorithms to determine on to which physical resources (servers) network functions are \textit{placed}, and be able to move functions from one server to another for such objectives as load balancing, energy saving, recovery from failures, etc. The task of placing functions is closely related to virtual network embedding \cite{VNEFischer13} and virtual data center embedding \cite{RabbaniMG13} and may therefore be formulated as an optimization problem, with a particular objective. Such an approach has been followed by \cite{BastaA2014, Moens14, Bagaa14, Mehraghdam14, BouetM15}. 

For example, Basta et. al \cite{BastaA2014} investigated the influence of virtualizing the S-GW and P-GW functions on the transport network load and data-plane delay. For these two functions, the authors showed differences in performance (of upto 8 times) when the functions were either fully virtualized and when their data and control planes were separated. The authors proposed a model for placing the functions in a way that minimizes the network load overheads introduced by the SDN control plane interactions. In addition to placement, Mehraghdam \cite{Mehraghdam14} proposes a model for formalizing the chaining of \acp{NF}. To this end, for each service deployment request, their approach constructs a VNFFG which is then mapped to the physical resources, considering that the network resources are limited and that functions have specific requirements. The mapping is formulated as a Mixed Integer Quadratically Constrained Program (MIQCP). The authors concluded that in order to obtain efficient use of resources, the placement of functions should be different according to the desired placement objective (i.e. remaining data rate, latency, number of used network nodes). Finally, Moens et. al \cite{Moens14} formulate the placement problem as an \ac{ILP} with an objective of allocating a service chain onto the physical network minimizing the number of servers used.

However, when formulated as an optimization problem, function placement and chaining would  reduce to a binary integer program, which is NP-Hard \cite{Schrijve86}, and hence intractable for big instances of the problem. This calls for heuristics such as those proposed in \cite{Xia15, Yoshida14, Hyeonseok14, Clayman14}. For example, Xia et. al \cite{Xia15} formulate the placement and chaining problem as binary integer programming (BIP), and propose a greedy heuristic to improve computational efficiency. The proposed greedy algorithm first sorts \acp{VNF} according their resource demand, and thereafter, \acp{VNF} with the highest resource demands are given priority for placement and chaining.

In addition, NFV systems should allow for one or a group of VNFs to be migrated to disparate physical servers. The physical servers may be in different InP domains, and hence use different tunneling addresses or be managed by different protocols. This does not only call for efficient algorithms to determine where the functions can be moved, but will also require comprehensive management of function and server states, as well as maintain communications. ViRUS  \cite{LWannerS14} allows the runtime system to switch between blocks of code that perform equivalent functionality at different \ac{QoS} levels when the system is under stress, while \cite{CerroniW14} presents a model that can be used to derive some performance indicators, such as the whole service downtime and the total migration time, so as to make function migration decisions.

%


Finally, to ensure scalable NFV implementations, functions should only be allocated the resources they need. Contrary to most current proof of concept implementations, it is not feasible to deploy a VM per subscriber or per function as the resulting VM footprint would be too high. This is because each VM is like a computer running its own operating system, and is meant to be isolated from other VMs and hence independent on a network level. This approach could become wasteful of resources for two reasons: (1) some of the functions such as DHCP in a CPE are so light that they would not justify a dedicated operating system on the scale of multiple functions per user, (2) some functions do not need to be strictly isolated from each other. Therefore, depending on the requirements of a given function, containers could be a more efficient way to use resources. Linux containers \cite{LXC} are an alternative to dedicated VMs in which a Docker \cite{Docker} may be used to achieve the automated resource isolation and namespacing which allows for partitioning of memory, network, processes etc. The use of containers avoids the overhead of starting and maintaining virtual machines since they do not require a complete duplication of an operating system. Using containers could lead to up to a 30\% savings in server costs to support the same number of virtual hosting points \cite{TNolleDocker}.

Moreover, even if given functions must utilize the same resources in a VM's operating system, it is possible to use scheduling techniques to allow the functions to share the resources. To this end, the proposals in \cite{MijumbiNFV15, Ferrer14, Escalona14} formulate the problem as a Resource Constrained Project Scheduling Problem (RCPSP) \cite{Brucker19993} and solve it using a job shop scheduling approach \cite{Blazewicz19961}. Specifically, Mijumbi et. al \cite{MijumbiNFV15} formulate an online \ac{VNF} mapping and scheduling problem and propose a set of greedy algorithms and a Tabu Search (TS) \cite{Glover97} heuristic for solving it. The greedy algorithms perform the mapping and scheduling of VNFs based on a greedy criterion such as available buffer capacity for the node or the processing time of a given VNF on the possible nodes, while the TS algorithm starts by creating an initial solution randomly, which is iteratively improved by searching for better solutions in its neighborhood. 

In addition, existing scheduling tools such as Google's Borg \cite{VermaPKOTW15} and Apache Mesos \cite{Hindman11} may be considered for scheduling of \acp{VNF}. Borg uses task-packing, over-commitment, and machine sharing with process-level performance isolation to run multiple jobs, from many applications, across a number of clusters. Users of the Borg system submit jobs consisting of one or several tasks that are run from the same executable. The scheduler in Borg monitors queues and schedules jobs considering the resources available on individual machines. The jobs may have requirements such as CPU and OS. However, unlike the functions in NFV, the tasks in Borg are run directly on hardware not in a virtualized environment. In addition, while Borg may have the scalability (cells usually contain 10K servers) that would be required in an NFV environment, it would have to be improved to meet carrier class requirements. For example, unlike the functions that make up a service in NFV, the tasks considered in Borg do not have ordering requirements. Finally, a task start up latency of 25s, and the 4 nines ($99.99\%$) availability that Borg is able to give may need to be enhanced for NFV.

Therefore, it can be observed that there are still many open areas with regard to how physical resources are shared among the \acp{VNF}. First of all, the results in each of the above areas may still be improved. In particular, the efficiency and applicability of containers needs to be studied more, just like there is need to study and propose more efficient function scheduling algorithms. In addition, given the dynamic requirements of NFV, there is need for resource allocation proposals that are able to find solutions online, consider multi-domain and distributed \acp{VNF} \cite{Ahmed15, Rosa15}, network survivability \cite{RabbaniZB14}, dynamic resource management \cite {MijuRash14} etc.

\subsection{Security, Privacy and Trust}

Despite the enormous potential of cloud computing, consumer uncertainty and concern regarding issues of privacy, security and trust remain a major barrier to the switch to cloud models \cite{PearsomYee13}. Therefore, cloud privacy issues will be among the key concerns for TSPs if they have to move to public clouds. Because the functions to be virtualized represent subscriber services, personally identifiable information may be transferred to the cloud. This will present unique challenges especially as the functions will be distributed, making it hard to know where this data is and who has access to it. In the case where the functions are deployed in third party clouds, users and Telecom service providers would not have access to the physical security system of data centers. Even if the service providers do specify their privacy and security requirements, it may still be hard to ensure that they are fully respected.

Emphasizing its importance, ETSI constituted a security expert group to focus on this concern. The group started by identifying potential security vulnerabilities of NFV and establishing whether they are new problems, or just existing problems in different guises \cite{ETSINFV011X}. The evaluation confirmed that indeed NFV creates new security concerns as shown in Table \ref{secthe}. After identifying the possible threats, the group proposed some solutions. In particular, they have provided a security and trust guidance that is unique to NFV development, architecture and operation \cite{ETSINFV011}. However, this does not consist of prescriptive requirements or specific implementation details.
\begin{table}[t]
\caption{Potential Security Threats in NFV \cite{ETSINFV011X}}
\label{secthe}
\small
\renewcommand{\arraystretch}{1.4}
\centering
\rowcolors{2}{gray!25}{white}

\begin{tabular}{ L{7.9cm}} \hline
Security Threat \\ 
Topology Validation \& Enforcement \\ 
Availability of Management Support Infrastructure \\ 
Secured Boot \\ 
Secure Crash \\ 
Performance Isolation \\ 
User/Tenant Authentication, Authorization and Accounting \\ 
Authenticated Time Service \\ 
Private Keys within Cloned Images \\ 
Back-Doors via Virtualized Test \& Monitoring Functions \\ 
Multi-Administrator Isolation \\ \hline  

\end{tabular}
\end{table}

However, it was noted that while solutions for these threats are available, there are currently no processes to take advantage of these solutions and, once in place, they will add procedural complexity \cite{Cui14, ETSINFV011X}. Moreover, for some of the threats (such as topology validation, network performance isolation and multi-administrator isolation), the group determined that solutions are not yet available \cite{Cui14}. As NFV gets deployed and more important functions virtualized, we can expect it to attract even more security and privacy threats. More than ever, there will be threats based on data interception (whether lawful or otherwise). Therefore, security, privacy and trust are other important research directions in NFV.

\subsection{Modeling of Resources, Functions and Services}

NFV's potential is based on its ability to deliver high levels of automation and flexibility. However, the resources and functions in NFV will be provided by different entities. Therefore, the availability of well understood, open and standardized descriptors for these multi-vendor resources, functions and services will be key to large-scale NFV deployments. Models should consider both initial deployment as well as lifecycle management - reconfiguration. As part of the \ac{MANO} specification \cite{ETSINFV009}, the ETSI provided a possible set of models that may be useful in NFV. These include OVF, TOSCA, YANG and SID. OVF was introduced in section \ref{dtmfovfref}. In what follows, we introduce the other three models.

\begin{table*}[!htbp]
\caption{Summary of Choice of Information and Data Models for NFV}
\label{modelsum}
\centering
\includegraphics[width=18.2cm, height=8.30cm]{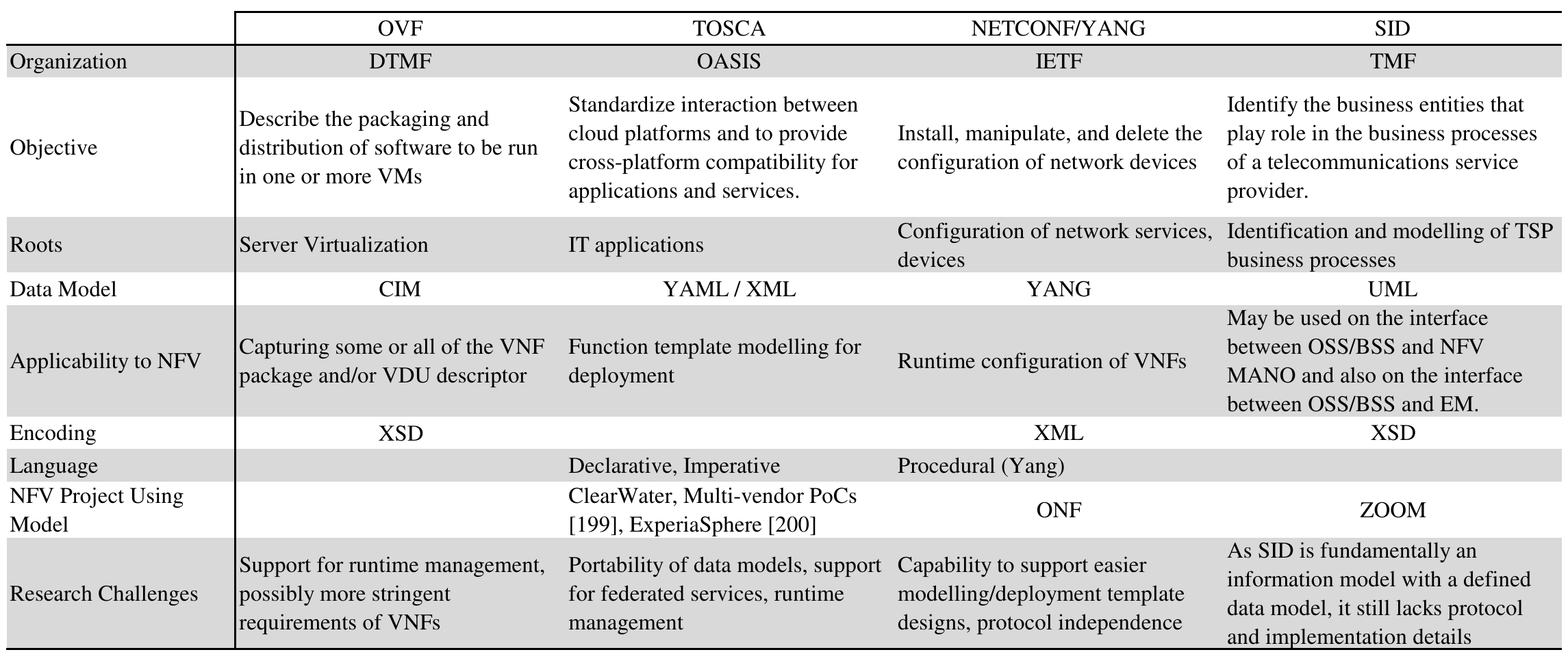}
\end{table*}
\subsubsection{Topology \& Orchestration Standard for Cloud Application (TOSCA)}
TOSCA \cite{TOSCA} is an OASIS standard language to describe a topology of cloud based web services, their components, relationships, and the processes that manage them. It describes what is needed to be preserved across service deployments in different environments to enable inter-operable deployment of cloud services and their management when the applications are ported over alternative cloud environments \cite{ETSINFV009}. TOSCA may be used for VNF definition, node monitoring and active policies like healing and scaling.

\subsubsection{NETCONF/YANG}

NETCONF \cite{Schonwalder10} is a protocol defined by the IETF to ``install, manipulate, and delete the configuration of network devices". NETCONF operations are realized on top of a Remote Procedure Call (RPC) \cite{Birrell1984} layer using an XML encoding and provide a basic set of operations to edit and query configuration on a network device. NETCONF is based on the YANG data modeling language. YANG is used to model both configuration and state data of network elements. Furthermore, YANG can be used to define the format of event notifications emitted by network elements and it allows data modelers to define the signature of remote procedure calls that can be invoked on network elements via the NETCONF protocol. 


\subsubsection{Information Framework (SID)}
SID \cite{TMFSID} is a component of TM Forum's Frameworx aimed at providing an information model and common vocabulary for all the information shared among things of interest (entities) to an enterprise such as customer, location and network element, and relationships (associations) between these entities, such as a network element is situated at a location. Entities are further characterized by facts (attributes) that describes them and their behavior (operations) that describe how the entities work. SID was originally based on Unified Modeling Language (UML) \cite{Rumbaugh04UML}, but was extended to include XML Schema Definition (XSD) representations.\\

\noindent \textbf{Discussion:} Table \ref{modelsum} summarizes the information and data modeling possibilities for NFV. All the models defined above have relatively wide adoption, and may therefore be considered for modeling of resources and functions in NFV. For example, to enable simple and scalable gradual deployment of VNFs and other NFV concepts, VNFs need to co-exist with traditional non NFV-based NFs. To provide an integration with existing OSS/BSS systems, end-to-end network services that include VNFs or VNF Forwarding Graphs may be able to be mapped to the SID service model \cite{ETSINFV009}.

However, these models were not initially developed with explicit considerations for some of the more specific requirements expected by NFV deployments and can therefore only be used as starting points and should continue to evolve for this purpose. For example, portability of data models and support for federated services have been identified \cite{Katsaros14, binz2013opentosca} as outstanding improvements for TOSCA. TOSCA also needs improvement to support run-time management of services. With regard to NETCONF/YANG, there is need to improve them to be able to cope with situations when multiple administrators (multi-domain environment) are present \cite{ParkJames14}. A lot of work is ongoing to extend some of the models for NFV. For example, SID has been extended using the ZOOM information model \cite{TMFSIDImpr15} to define four concepts (VirtualResource, NetworkFunction, NetworkService, and Graph) aimed at modeling NFV-based systems. In addition, The TOSCA TC recently formed a workgroup focused on creating a ``TOSCA Simple Profile for NFV."

As the models continue to improve, it may be important to have solutions that combine them so as to avoid some of their disadvantages. For example, A TOSCA template can install a virtual router, but it cannot subsequently create/modify/delete configuration on demand on the same router during run-time. Therefore, fulfilling VNF requirements requires more than TOSCA. 
In the same way, YANG is designed for writing machine readable schema, and is hence difficult to use for design of templates for initial service deployment. In this case, TOSCA may be combined with NETCONF/YANG where the file-based templates in TOSCA may be used for deploying VNFs on cloud infrastructure, while NETCONF can be used to provide a runtime API both for configuring VNFs after they have been installed, bringing VNFs to a state of operational readiness, and while they are running in the cloud, fulfilling the service requirements of a particular customer \cite{HeavyReading}.

\begin{table*}[!htbp]
\caption{Summary of State-of-the-art and Research Challenges}
\label{challenges}
\centering
\includegraphics[width=17.15cm, height=21.60cm]{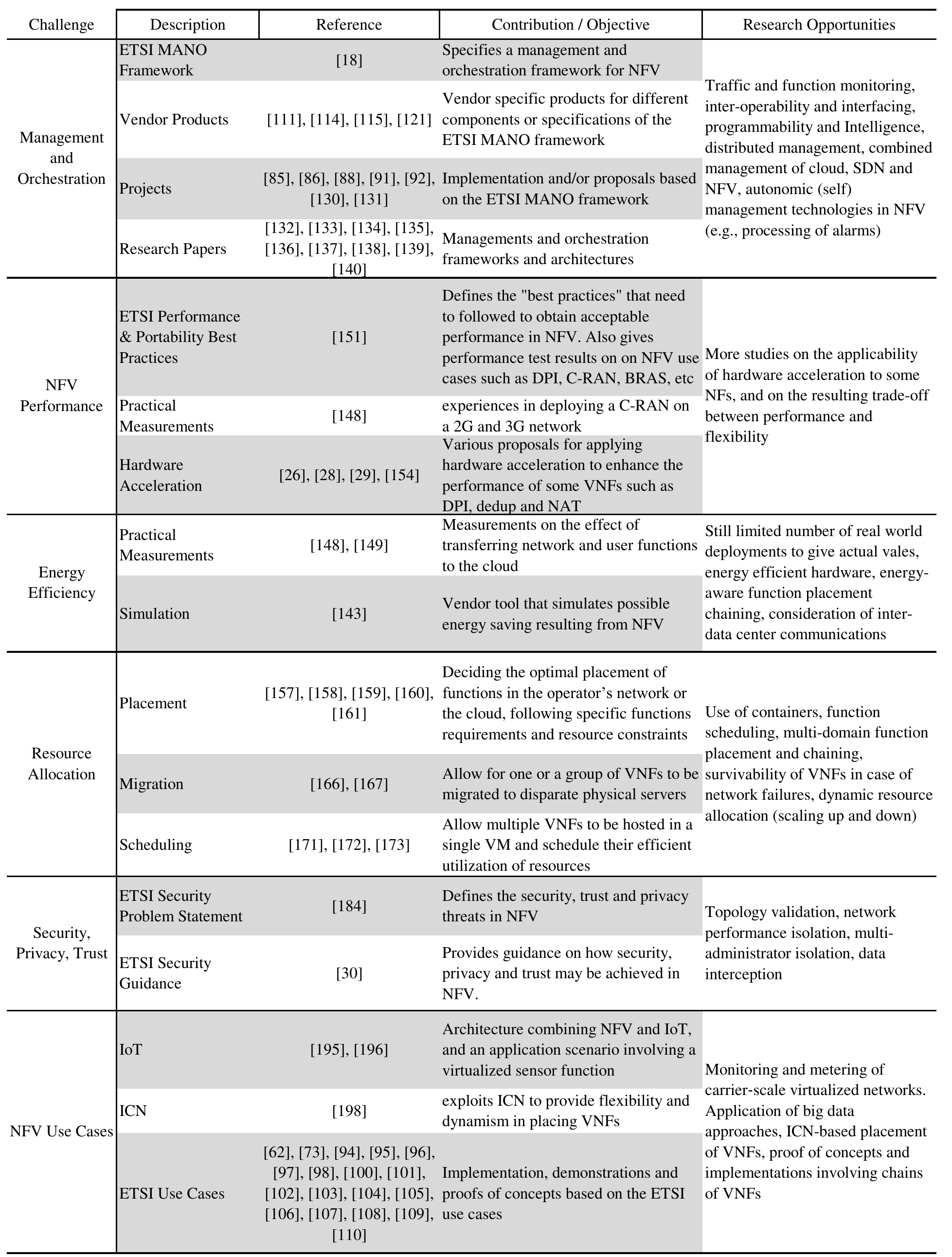}
\end{table*}

\subsection{Research Directions in Selected NFV Use Cases}

\subsubsection{The Internet of Things}

Like \ac{NFV}, the Internet of Things (IoT) \cite{Gubbi2013} paradigm has recently drawn a lot of industrial attention. The IoT is a network of physical objects or ``things" into which sensors with unique identifiers are embedded. Such sensors may collect and transfer various kinds of (big) data over a network without requiring human-to-human or human-to-computer interaction. Inevitably, by networking zillions of devices, the IoT will lead to networks of unbelievable scale and complexity with tremendous implications on network management. It will lead to security, scalability and resource management challenges in networks that should simultaneously transport, process and act on this data in real time.

NFV has been proposed as a key enabler of the IoT \cite{iotnfv, OmnesN15}. The idea in \cite{iotnfv} is to limit the functionality embedded in deployed sensors, and provide virtualized functions such as security, intelligence, computation and storage to the devices. These would take advantage of the scalable distribution capabilities of NFV as well as the configuration flexibility of SDN. On the other hand, Omnes et. al \cite{OmnesN15} propose multi-layered IoT architecture involving SDN and NFV, and illustrate how the proposed architecture is able to cope with some of the challenges in IoT.

However, there are serious questions on the management of big amounts of IoT-generated data with better network efficiency. It is therefore critical to study efficient ways of transporting (big) data over such sofwarized networks, and whether current cloud data management applications such as Hadoop and Cassandra would be able to support the real time requirements in such environments.

%
%
%
%
%
%
%
%

\subsubsection{Information-Centric Networking}
%
%

Motivated by the fact that the Internet is increasingly used for information dissemination rather than for pair-wise communication between end hosts, Information-Centric Networking (ICN) \cite{Xylomenos14} has emerged as a promising candidate for the architecture of the Future Internet. ICN addresses named data rather than named hosts. This way, content distribution is implemented directly into the network fabric rather than relying on the complicated mapping, availability, and security mechanisms currently used to map content to a single location.

The separation between information processing and forwarding in ICN is related to both the decoupling of functions from devices in NFV, and to the decoupling of control from data plane in SDN. While the relationship between NFV, SDN and cloud computing has already received some attention, that between NFV and ICN has not. Yet, ICN may be used in NFV to determine the best position to place network functions. For example, Arumaithurai et al. \cite{Arumaithurai14} propose a \textit{function-centric service chaining} (FCSC) approach which exploits ICN to provide flexibility and dynamism in placing VNFs.\\\\

\noindent \textbf{Summary:} In Table \ref{challenges}, we summarize the state-of-art in each of the identified research challenges, as well as specific open questions in each one of them. We have noted that despite the significant and rapidly increasing activity on NFV, there are still major gaps especially with regard to standardization that may slow down NFV deployment and undermine the possibilities to fulfill its anticipated business case. While the ETSI-defined reference architecture covers most of the aspects needed to operationalize NFV, current specifications are still too general to envelope all the essential pillars of required evolution such as inter-operability, legacy support, and management of both legacy and NFV-based systems.

For example, currently, different vendors depend on different languages to model resources and functions in NFV. TOSCA has been used in modeling services for a multi-vendor E2E proof of concept \cite{IntelBrocadeTel}, for ClearWater, and ExperiaSphere \cite{ExperiaSphere}. On the other hand, the descriptors in the HP NFV Director are not based on TOSCA, and ONF has chosen YANG as the modeling language. Similar examples can be given for vendor implementations of the NFVI, VNFs and MANO. This could result into inter-operability challenges where vendor-specific Command Line Interfaces (CLIs) require manual configuration or expensive integration by service providers themselves or systems integrators with their own proprietary tools and equipment-specific adapters. Therefore, though there are many options for modeling of functions and resources, the techniques remain generally in their infancy.

With regard to performance, most current PoCs are based on a rather limited list of use cases proposed by the ETSI. While these PoCs are important to prove technical principles unique to NFV, they do not give a complete view of performance and benefits for a wide range of end-to-end services. Finally, research on possible enablers of NFV such as ICN, and on the application areas such as IoT are still largely unexplored.

\section{Conclusion}\label{conclusion}

Due to user demands for real-time, on-demand, online, inexpensive, short-lived services, TSPs have been forced to look for new ways of delivering these services in ways that are agile, and with OPEX and CAPEX savings. NFV has emerged as a possible approach to make network equipment more open, and hence allow TSPs to become more flexible, faster at service innovations and reduce operation \& maintenance (O\&M) costs. It is clear that NFV, together with the closely related and complementary fields of SDN and cloud computing may be big parts of future telecommunication service provision. 

In this paper, we introduced NFV, described its architecture as defined by ETSI, proposed a reference business model, and explored important design considerations. We then compared NFV with closely related fields, SDN and cloud computing, discussing current research for combining them. We have also presented major specification and standardization  efforts, research projects, commercial products and early NFV proof of concept implementations. Finally, we discussed the key research areas that will be pivotal to the success of NFV as well as to its application to ICN and IoT, and summarized the findings of the survey. We believe that before these areas are explored, TSPs who deploy NFV may end up being reliant on vendor-proprietary solutions to solve these gaps, which would be against the original objective of NFV.

We have noted that many current NFV solutions, especially from the industry, have been mainly about pooling vendor specific resources hosted in a cloud rather than real support for flexibility, inter-operability, integrated management, orchestration and service automation all of which are core requirements for NFV. It is expected that such implementations will continue to increase before NFV gets completely standardized. As NFV moves from labs and PoCs to trials and commercial deployments, vendors are investing significant resources to develop these NFV solutions. It is therefore urgent for specification and standardization bodies to complete specifications before it becomes too late for the standards to change or influence what has already been deployed.

\section*{Acknowledgment}
The authors are indebted to the Editor-in-Chief for coordinating the review process, and to the anonymous reviewers for their insightful comments and suggestions. This work was partly funded by FLAMINGO, a Network of Excellence project (318488) supported by the European Commission under its Seventh Framework Programme, and project TEC2012-38574-C02-02 from Ministerio de Economia y Competitividad.

\bstctlcite{IEEEexample:BSTcontrol}
\bibliographystyle{IEEEtran}
\bibliography{IEEEabrv,nfvbiblio}

\begin{IEEEbiography}
[{\includegraphics[width=1in,height=1.25in,clip,keepaspectratio]{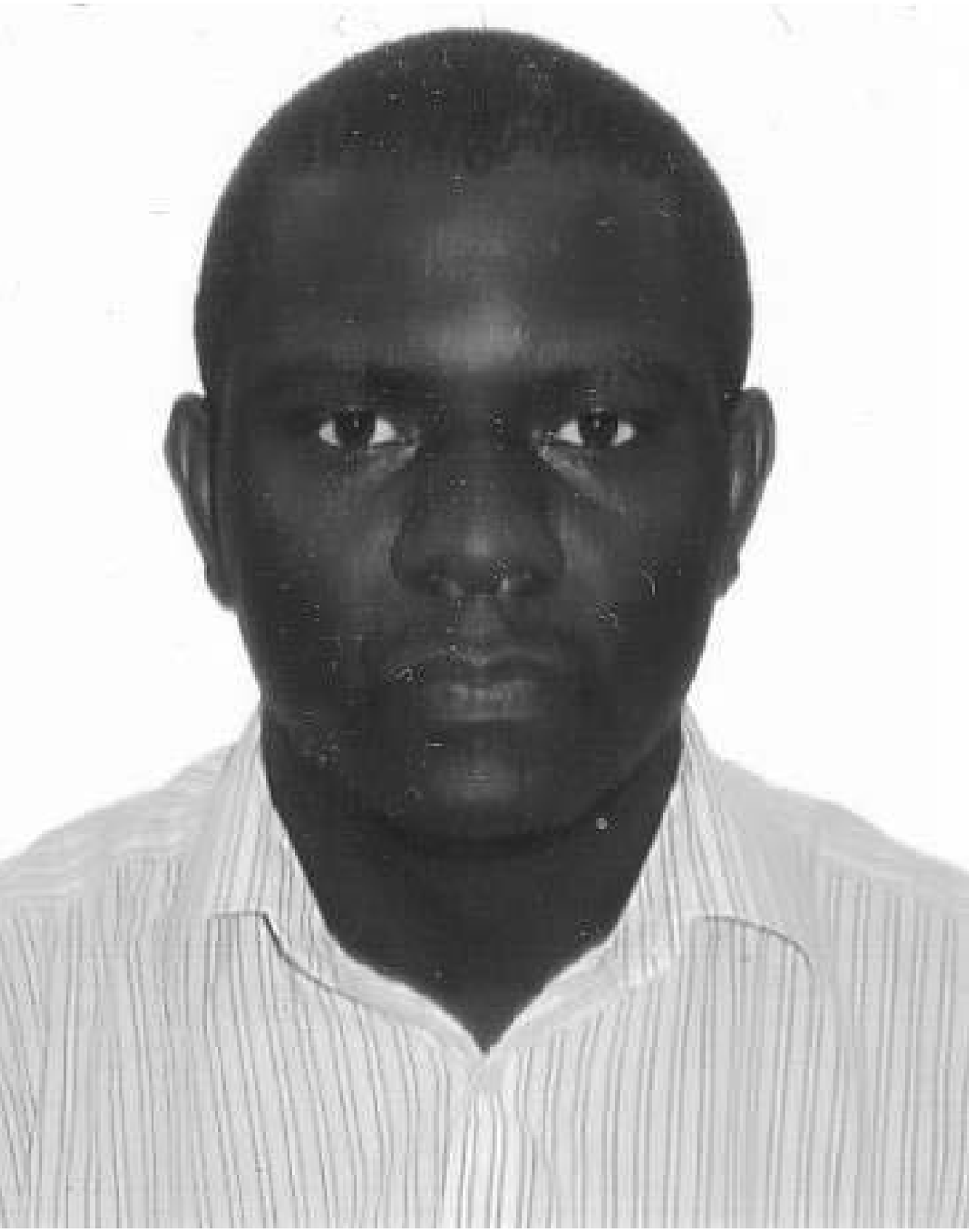}}]{Rashid Mijumbi}
obtained a degree in electrical engineering from Makerere University, Uganda in 2009, and a PhD in telecommunications engineering from the Universitat Polit\`{e}cnica de Catalunya (UPC), Spain in 2014. He is currently a Postdoctoral Researcher in the Network Engineering Department at the UPC. His research interests are in autonomic management of networks and services. Current focus is on management of resources for virtualized networks and functions, cloud computing and software defined networks.
\end{IEEEbiography}

\begin{IEEEbiography}
[{\includegraphics[width=1in,height=1.25in,clip]{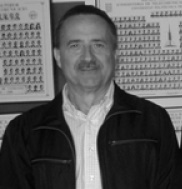}}]{Joan Serrat} received a degree of telecommunication engineering in 1977 and a PhD in the same field in 1983, both from the Universitat Polit\`{e}cnica de Catalunya (UPC). Currently, he
is a full professor at UPC where he has been involved in several collaborative projects with different European
research groups, both through bilateral agreements or through participation in European funded projects. His
topics of interest are in the field of autonomic networking and service and network management. Currently, he is the contact point of the TM Forum at UPC. 
\end{IEEEbiography}

\begin{IEEEbiography}
[{\includegraphics[width=1in,height=1.25in,clip]{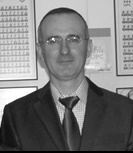}}]{Juan-Luis Gorricho}
received a telecommunication engineering degree in 1993, and a Ph.D. degree in
1998, both from the UPC. He is currently an associate professor at the UPC. His recent research interests are in applying artificial intelligence to ubiquitous computing and network management; with special interest on using smartphones to achieve the recognition of user activities and locations; and applying linear programming and reinforcement learning to resource management in virtualized networks and functions.
\end{IEEEbiography}

\begin{IEEEbiography}
[{\includegraphics[width=1in,height=1.25in,clip,keepaspectratio]{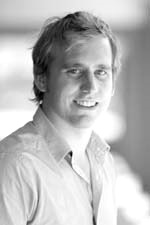}}]{Niels Bouten}
obtained a masters degree in computer science from Ghent University, Belgium, in June 2011. In August 2011, he joined the Department of Information Technology at Ghent University, where he is active as a Ph.D. student. His main research interests are the application of autonomic network management approaches in multimedia delivery. The focus of this research is mainly on the end-to-end Quality of Experience optimization, ranging from the design of a single autonomic control loop to the federated management of these distributed loops.
\end{IEEEbiography}

\begin{IEEEbiography}
[{\includegraphics[width=1in,height=1.25in,clip,keepaspectratio]{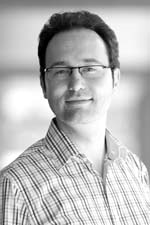}}]{Filip De Turck} is a professor at the Department of Information Technology of Ghent University and iMinds in Belgium, where he leads the network and service management research group. His main research interests include scalable software architectures for network and service management, design and performance evaluation of novel QoE-aware multimedia delivery systems. He served as TPC chair of the IEEE/IFIP Network Operations and Management Symposium (NOMS 2012) and the IFIP/IEEE Integrated Network Management Symposium (IM 2013). He is associate editor of the Journal on Network and System Management, the International Journal of Network Management and IEEE Transactions on Network and Service Management.
\end{IEEEbiography}

\begin{IEEEbiography}
[{\includegraphics[width=1in,height=1.25in,clip,keepaspectratio]{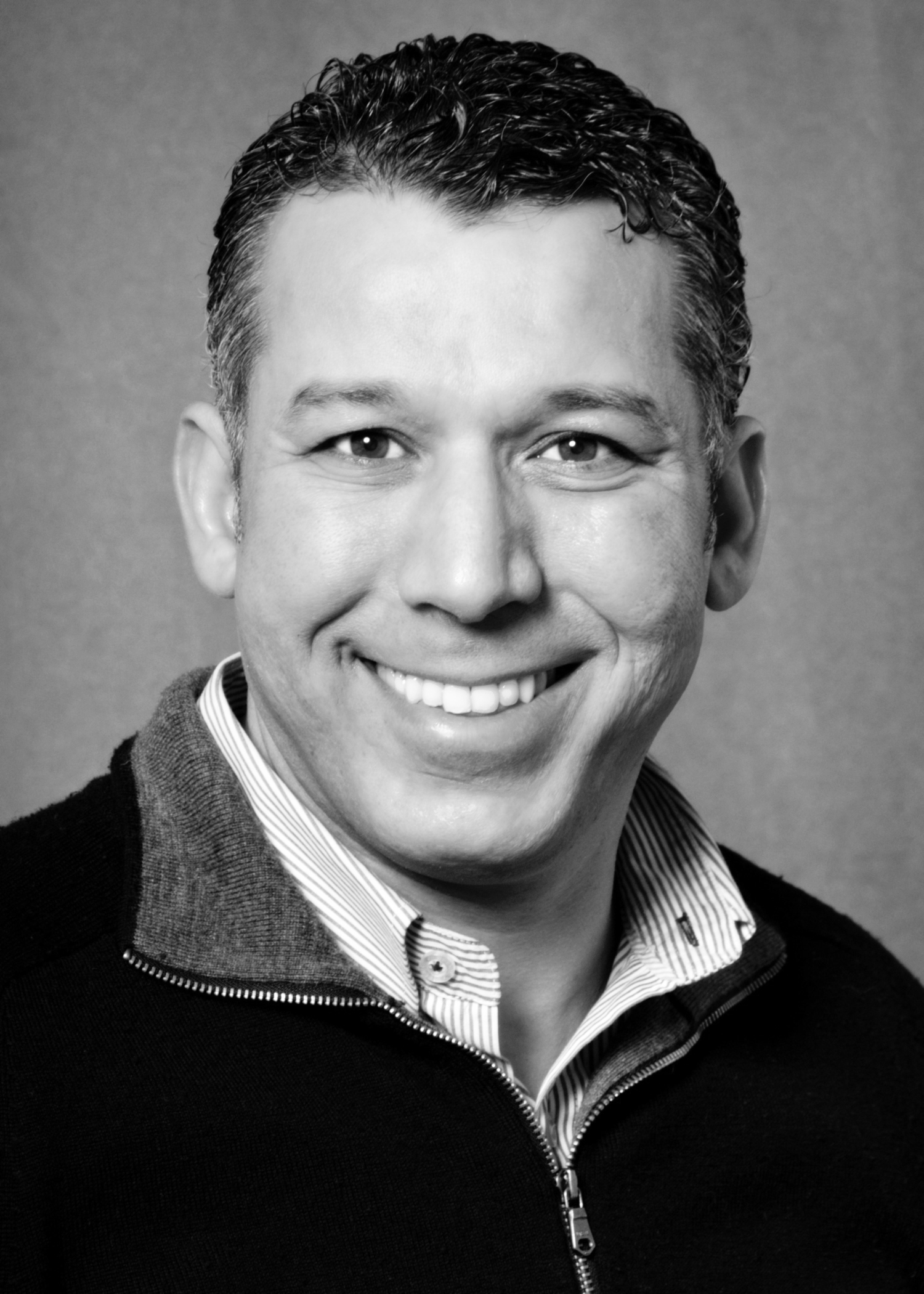}}]{Raouf Boutaba} received the MSc and PhD degrees in computer science from the Universit\'{e} de Pierre et Marie Curie, Paris, France, in 1990 and 1994, respectively. He is currently a full professor of computer science at the University of Waterloo, Waterloo, ON,  Canada, and a distinguished visiting professor at the Pohang University of Science and Technology (POSTECH), Korea. His research interests include network, resource and service management in wired and wireless networks. He has received several best paper awards and other recognitions such as the Premier's Research Excellence Award, the IEEE Hal Sobol Award in 2007, the Fred W. Ellersick Prize in 2008, the Joe LociCero and the Dan Stokesbury awards in 2009, and the Salah Aidarous Award in 2012. He is a fellow of the IEEE and the Engineering Institute of Canada.
\end{IEEEbiography}

\end{document}